\documentclass[prodmode,acmtaco]{acmsmall} 

\usepackage[linesnumbered, ruled]{algorithm2e}

\SetAlFnt{\small}
\SetAlCapFnt{\small}
\SetAlCapNameFnt{\small}
\SetAlCapHSkip{0pt}
\usepackage{amsmath}
\usepackage{multirow}
\usepackage{subfig}
\usepackage{comment}
\usepackage{tipa}
\usepackage{tikz}
\usepackage{booktabs}
\usepackage{color,soul}
\newcommand{\redHL}[1]{\textcolor{black}{#1}}
\newcommand{\ignore}[1]{}

\begin{document}

\markboth{}{On Dynamic Precision Scaling}

\title{On Dynamic Precision Scaling 
}
\author{SERIF YESIL*
\affil{Bilkent University}
ISMAIL AKTURK*
\affil{University of Minnesota, Twin Cities}
ULYA R. KARPUZCU
\affil{University of Minnesota, Twin Cities}
}

\begin{abstract}
Based on the observation that application phases exhibit varying degrees of
sensitivity to noise (i.e., accuracy loss) in computation during execution, this
paper explores how Dynamic Precision Scaling (DPS) can maximize power efficiency
by tailoring the precision of computation adaptively to temporal changes in
algorithmic noise tolerance.  DPS can decrease the arithmetic precision of
noise-tolerant phases to result in power savings at the same operating speed (or
faster execution within the same power budget), while keeping the overall loss
in accuracy due to precision reduction bounded.

\end{abstract}



\begin{bottomstuff}
*Serif Yesil is now at University of Illinois, Urbana-Champaign; Ismail Akturk, at University of Missouri, Columbia.
\end{bottomstuff}

\maketitle

\section{Introduction}
\label{sec:intro}

Practical stagnation in voltage scaling renders an increasing chip power density
(power per area) over technology generations.  At the same time, cooling and
power delivery limitations prevent a proportional expansion of the chip power
budget.  The only way to sustain performance improvement in this environment is
enhancing power efficiency, i.e., performance gain per unit power
consumed~\cite{Horow}.  {\em Approximate computing} is a promising paradigm
which can enhance power efficiency by trading computation accuracy for
performance or power, depending on the tolerance of algorithms to noise (i.e.,
accuracy loss) in computation.  The intrinsic noise tolerance of the emerging
R(ecognition), M(ining), and S(ynthesis) applications~\cite{rms} -- which
process massive but noisy input data by probabilistic algorithms (often
featuring iterative refinement) -- makes them particularly suitable to
approximate computing. 

Approximate computing by precision reduction represents a heavily explored area
of research~\cite{yeh2007,discApprox,enerJ}.  In this paper, we explore
approximation by {\em adaptive} precision reduction, a relatively less explored
area.  Specifically, based on the observation that applications exhibit varying
degrees of sensitivity to noise during computation, we explore {\em how
tailoring the precision of computation adaptively to temporal changes in
algorithmic noise tolerance can maximize power efficiency}. 

Due to its analogy to Dynamic Voltage and Frequency Scaling (DVFS), we refer to
this paradigm as {\em Dynamic Precision Scaling} (DPS).
Recall that DVFS can maximize power efficiency by tracking temporal changes in the
performance demand of the workload (which in turn evolves as a function of
temporal changes in computation and memory access characteristics) and by
changing the operating point (i.e., the operating voltage and frequency)
accordingly. 
Similar to DVFS, DPS tracks temporal changes in workload characteristics.
However, in maximizing power efficiency, contrary to DVFS, DPS exploits temporal
changes in noise tolerance, and adaptively decreases the arithmetic precision of
noise-tolerant phases to obtain power savings at the same operating speed (or
faster execution within the same power budget) while keeping the overall loss in
accuracy due to precision reduction bounded.  \redHL{Conceptually,} for less noise-tolerant phases,
DPS can adjust the arithmetic precision on the fly to prevent excessive loss in
computation accuracy.

\redHL{In this paper, we conduct a limit study 
in order to quantify the power efficiency potential of DPS. To this end, we
devise a proof-of-concept implementation of the DPS concept.
Specifically,} we first develop
a workload analyzer which can identify application phases of varying noise
tolerance. Then, using the outcome of this workload analyzer, we design two
heuristic DPS policies.  In the following, 
Section~\ref{sec:dps} introduces DPS basics and the two \redHL{proof-of-concept} policies;
Sections~\ref{sec:setup} and~\ref{sec:eval} provide the evaluation;
Section~\ref{sec:rel} covers related work; 
and Section~\ref{sec:conc} summarizes our findings.

\section{Dynamic Precision Scaling (DPS)}
\label{sec:dps}
\subsection{Dynamic Precision Scaling: Basics}

To be able to tune the arithmetic precision of computation on the fly, any
practical DPS implementation has
to monitor fine grain temporal changes in the noise tolerance of the
workload.  
Noise tolerance of emerging  RMS
applications stems from (i) algorithms which are mostly probabilistic and often
utilize iterative refinement; (ii) inputs which contain a very large number
of noisy (and often redundant) elements.  Due to (i), it is barely possible to
differentiate noise tolerant phases from others without understanding program
semantics.  
On the other hand, (ii) renders profiling-based identification of noise tolerant
phases inevitable, as explored in~\cite{rumba,debugApprox}. 

\redHL{
Based on these observations, we envision a practical DPS   
implementation to comprise three basic modules:
an {\em
offline profiler}, a {\em runtime monitor}, and an {\em accuracy controller}.
The differences in the design of these three modules give rise to different
points in the DPS design space. 
The {\em offline profiler} 
not only identifies but also demarcates noise tolerant application phases, such that 
the {\em runtime monitor} can detect on the fly which phases of the
application are more noise tolerant and which are less. 
Finally, the {\em 
accuracy controller} processes the output of the {\em runtime
monitor} to adjust the arithmetic precision on the fly. Similar to DVFS
controllers, the {\em accuracy controller} is in charge of scheduling decisions
for timely precision scaling.
}

\redHL{The underlying hardware architecture can harvest power efficiency from
  DPS in numerous ways.  
For example, arithmetic units of reconfigurable precision represent a good match for DPS. 
Narrow operand widths under lower precision arithmetic
can result in higher power efficiency due to the increased processing speed along
with 
power
savings~\cite{rutenbar}.   However, in mapping (more)
noise-tolerant phases to arithmetic units of reduced precision, particularly under
fine-(temporal)-grain DPS, the {\em accuracy controller} has to carefully budget for the
power and performance overhead of the scheduling decisions.  
}

\redHL{In the following, we will detail a proof-of-concept DPS implementation
  which features 
an {\em
offline profiler} along with a hypothetical {\em runtime monitor} and an {\em accuracy controller}.
This implementation primarily serves automated
design space exploration with limited user input which enables the application
programmer and system designer to exploit 
temporal variations in the noise-tolerant 
phases of
computation.}

\subsection{A Proof-of-Concept DPS Implementation}
\label{sec:policy}

\redHL{Without loss of generality, we confine the proof-of-concept DPS
implementation to the floating point datapath. However, the DPS concept
generally applies to the integer datapath, as well, where the main complication
comes from identification, and hence, exclusion of memory address calculations
(i.e., pointer arithmetic) from approximation.}

We reduce
precision  by omitting a subset of less significant bits of the fraction:
According to the IEEE 754 standard, a single (double) precision floating point
number occupies an 32(64)-bit register with one bit allocated for {\em sign}, 8
(11) bits for {\em exponent}, and 23 (52) bits for {fraction}, i.e., {\em
mantissa}, respectively. A single precision floating point number, e.g.,
corresponds to $(-1)^{sign}\times2^{exponent-127}\times1.mantissa$. 


\begin{algorithm}[t]
\SetAlgoNoLine
\KwIn{targetAccLoss, \#bits, \#dynamicCalls}
\KwIn {AccLossS0[1...\#dynamicCalls][1..\#bits]}
\KwIn {AccLossS1[1...\#dynamicCalls][1..\#bits]}
\KwOut{\#omittedBits[1...\#dynamicCalls]}

        \For {i$\gets$1...\#dynamicCalls
        }{
             $targetBit \gets 0$ \;
             $cummAccLoss \gets 0$ \;
                \While {cummAccLoss$<$targetAccLoss \& targetBit$\leq$\#bits} {
                
                        \eIf {AccLossS0[i][targetBit] and
						AccLossS1[i][targetBit] are valid }{
                                $err \gets \max (AccLossS0[i][targetBit],AccLossS1[i][targetBit])$ \;
                                $cummAccLoss+\gets err$ \;
                         }
                         {
				$break$\;
			 }
                        $targetBit++$\;
                }
                $\#omittedBits[i] \gets targetBit-1$\;
        }
\caption{\normalsize Basic DPS Policy}\label{alg:dps_algo1}
\end{algorithm}


\begin{algorithm}[t]
\SetAlgoNoLine
\KwIn{targetAccLoss, \#bits, \#dynamicCalls}
\KwIn {AccLossS0[1...\#dynamicCalls][1..\#bits]}
\KwIn {AccLossS1[1...\#dynamicCalls][1..\#bits]}
\KwOut{\#omittedBits[1...\#dynamicCalls]}

        \For {i$\gets$1...\hl{(\#dynamicCalls-1)}
        }{
		$targetBit \gets 0$ \;
                $cummAccLoss \gets 0$ \;
                \hl{cummAccLoss\_next $\gets$ 0}\;

		\While {cummAccLoss$<$targetAccLoss \& targetBit$\leq$\#bits 
				\newline \hl{ \& cummAccLoss\_next$<$targetAccLoss} 
                }{
                        \eIf {AccLossS0[i][targetBit], AccLossS1[i][targetBit],
                        \newline \hl{AccLossS0[i+1][targetBit]},
						\hl{AccLossS1[i+1][targetBit]} are valid
                        }{
                                $err \gets \max (AccLossS0[i][targetBit],AccLossS1[i][targetBit])$ \;
                                \hl{err\_next $\gets \max$ (AccLossS0[i+1][targetBit], AccLossS1[i+1][targetBit])} \;
                                $cummAccLoss+\gets err$ \;  
                                \hl{$cummAccLoss\_next+\gets err\_next$} \; 
                         }{
                                $break$\;
                        }
                         $targetBit++$\;
                }
                $\#omittedBits[i] \gets targetBit-1$\;
        }

\caption{Dependency-Aware DPS Policy: DPS+}\label{alg:dps_algo2}
\end{algorithm}

\redHL{The proof-of-concept implementation} captures temporal changes in application's noise tolerance by tracking
(dynamic) calls to floating point heavy functions within the course of
execution.
Dynamic function calls, i.e., different invocations of a given
(static) function, are dispersed in time within the course of execution,
and therefore can reflect temporal changes in noise tolerance.
\redHL{Without loss of generality, DPS can track temporal changes in
  application's noise tolerance at various granularities (such as instruction or
  basic block), giving rise to different implementations. The proof-of-concept
  design works at function granularity, and uses dynamic function calls to
  capture the notion of time. In other words, the proof-of-concept
  design employs noise tolerant
dynamic calls as proxies for noise tolerant phases of the application. 
  In this case, the question becomes {\em how to identify noise-tolerant dynamic
  function calls}. 
} 

\redHL{
To this end, the {\em offline profiler} module in the proof-of-concept implementation
uses a two-step approach: The first step
involves statistical fault injection; the second step,  post-processing of
statistical fault injection results. 
At the first stage, for each dynamic invocation of each floating point heavy function, we
corrupt one mantissa bit at a time and record the corresponding accuracy loss at
the application output. We repeat this experiment for all mantissa bits, by
injecting both stuck-at-0 and stuck-at-1 faults. We corrupt all floating point
variables in the function, in the same direction.  
Recall that the proof-of-concept
implementation uses noise tolerant
dynamic calls as proxies for noise tolerant phases of the application. The
accuracy loss observed in the end result per fault injection experiment serves
as a proxy for the degree of noise tolerance of each dynamic call.
The {\em offline profiler} also needs to communicate this information to
the runtime.}

\redHL{The post-processing step can rely on different policies.
We first devise a basic DPS policy 
following Algorithm~\ref{alg:dps_algo1}: The key inputs of this algorithm are {\em targetAccLoss},
the maximum accuracy loss in the end result the application can tolerate; and
the outcome of the first step of {\em offline profiling}, namely the accuracy loss observed in the
end result after injecting stuck-at-0 and stuck-at-1 faults in each mantissa
bit of each dynamic invocation of a floating point heavy function.  {\em
\#bits} specifies the number of mantissa bits 
subject to fault injection,
and {\em \#dynamicCalls}, the number of  
dynamic (floating point heavy) function calls (which may cover different static
functions). We keep the  
fault injection information in two separate ({\em
\#dynamicCalls}$\times${\em \#bits}) matrices for stuck-at-0 ({\em AccLossS0})
and stuck-at-1 faults ({\em AccLossS1}).  The output is the total number of
(consecutive) mantissa bits (starting from the least significant) we can omit
while the corresponding accuracy loss in end result remains lower than {\em
targetAccLoss}, on a per dynamic call basis: {\em \#omittedBits}.
}

Each step of the algorithm processes a different dynamic call (line 1). Starting
from the least significant bit, we check the accuracy loss in the end result
under the corruption of each mantissa bit (i.e., {\em targetBit}): If the
corresponding {\em AccLossS0(1)} entries are valid, i.e., the fault injection
experiment did not result in Inf or NaN (line 5), we extract the maximum of
accuracy loss under stuck-at-0 and stuck-at-1 (line 6).  This basic DPS policy
accumulates this maximum accuracy loss in the end result due to the corruption
of each mantissa bit in isolation (in {\em cummAccLoss} from line 7), as we
consider more mantissa bits for omission. {\em cummAccLoss} serves as a running
estimate for the actual accuracy loss in the end result. Accordingly, the policy
keeps processing higher-order mantissa bits for omission as long as  {\em
cummAccLoss} remains below {\em targetAccLoss} (line 4). 

A {\em runtime monitor} can then use the output of the basic DPS policy captured by
Algorithm~\ref{alg:dps_algo1}, {\em \#omittedBits}, to tune the precision of
each dynamic function call on the fly. Algorithm~\ref{alg:dps_algo1}'s main
bottleneck, however,  is {\em cummAccLoss}, the estimate of cumulative accuracy loss in the
end result of the application if we omit multiple mantissa bits (line 7). This
is because the actual (runtime) impact of each omitted mantissa bit on the
accuracy loss in the end result may not always be additive. Therefore, in the
worst case, if we omit multiple mantissa bits following
Algorithm~\ref{alg:dps_algo1} -- as captured by  {\em \#omittedBits} -- we may
eventually observe a higher accuracy loss in the end result than {\em
targetAccLoss}.
A refined version of the basic DPS policy, DPS+, can mitigate this, as depicted
in Algorithm~\ref{alg:dps_algo2}~\footnote{Recall that the for loop iterates until {\em
  \#dynamicCalls}-1 in this case. At the for loop exit, we cover the very last
dynamic call, not shown in the listing for brevity.} (with the difference between the two algorithms
highlighted).

Both algorithms process all dynamic function calls within the course of
execution; the order of the dynamic calls in {\em AccLossS0(1)} and {\em
\#omittedBits} arrays reflect their execution order in the {\em offline profiling} run. These
dynamic calls may cover more than one static (floating point heavy) function.
For the representative set of RMS benchmarks we experiment with
(Section~\ref{sec:setup}), we observe that dynamic calls following each other in
dynamic control flow are often also data dependent.
Algorithm~\ref{alg:dps_algo2} leverages this insight by limiting the precision
reduction of a dynamic function to the precision reduction of its follower
function in execution (and processing) order. In this manner, we enforce that
the (reduced) precision of a producer's output data matches (i.e., does not
exceed) the maximum acceptable precision of the input data of its (immediate)
consumer. 

Algorithm~\ref{alg:dps_algo2} still cannot provide mathematical guarantees (as
data dependent functions are not always executed back to back), but
can effectively enforce runtime accuracy loss (in the end result) to remain
below {\em targetAccLoss}. 
Algorithm~\ref{alg:dps_algo2} can further be refined by tracking actual call
graphs, similar to~\cite{fpCompiler}, of data
dependent dynamic functions for precision matching.

\section{Evaluation Setup}
\label{sec:setup}
\begin{table}[ht]
\centering
\caption{List of benchmarks used.}
\label{tab:benchmarks_detailed}
\scalebox{0.9} {
\begin{tabular}{|l|l|l|l|}
\hline
{\textbf{Benchmark}} & {\textbf{Description}} & {\textbf{Input Dataset}} \\ 
\hline
\textbf{Blackscholes (BS)}~\cite{parsec}	& PDE solver			& simsmall\\ 
\hline
\textbf{Fluidanimate (FA)}~\cite{parsec}	& n-body simulation     & simsmall \\ 
\hline
\textbf{Hotspot (HS)}~\cite{rodinia}		& Thermal simulation	& 64x64 grid    \\ 
\hline
\textbf{Particlefilter (PF)}~\cite{rodinia}	& Medical imaging		& 128x128 10 timepoints   \\ 
\hline
\textbf{Pagerank (PR)}~\cite{gapbs}			& Graph processing      & gnutella04   \\ 
			&       & \cite{snapnets}   \\ 
\hline
\end{tabular}
}
\end{table}

\redHL{Throughout the evaluation, we will refer to the proof-of-concept
implementation as DPS(+)}.  

\subsection{Benchmarks}
To quantify the power efficiency potential of DPS, we deploy a representative,
relatively floating point heavy, set of RMS applications from
PARSEC~\cite{parsec}, Rodinia~\cite{rodinia} and Gapbs~\cite{gapbs} suites, as captured by
Table~\ref{tab:benchmarks_detailed}. 
We deploy default input data sets except Pagerank (PR). The inputs for PR come
from a well-known graph database~\cite{snapnets}.

{\em Blackscholes (BS)} calculates stock option prices
solving Partial Differential Equations (PDE). \redHL{While the application
domain is arguably not suitable for approximation, we included the benchmark as
a representative PDE solver.}
The region of interest, ROI (where the actual computation takes place), comprises
the function \textit{BlkSchlsEqEuroNoDiv}.  
Each function call calculates the price of a different stock
option, hence errors in a call can only affect 
the corresponding stock price, but not other calls.
{\em Fluidanimate (FA)} represents n-body simulation, where bodies
correspond to fluid particles. 
The ROI comprises four relatively floating point heavy functions:
\textit{RebuildGrid}, \textit{ComputeForces}, \textit{ProcessCollisions},
\textit{AdvanceParticles}. 
{\em Hotspot (HS)} is an architectural thermal simulator. 
The ROI comprises a kernel to compute
the temporal temperature difference (i.e., delta) 
at each sampled 
location in a processor chip, 
encapsulated 
in \textit{find\_delta} function. 
{\em Particlefilter (PS)} is a medical imaging application to 
track
an object in an image.  The ROI comprises
\textit{particleFilter} function. The main work of this function is carried
out in a loop that iterates over frames. Every iteration calls five relatively
floating point heavy functions subject to DPS:
\textit{apply\_motion\_model}, \textit{particle\_filter\_likelihood},
\textit{update\_weights}, \textit{normalize\_weights}, and \textit{calc\_U}.
{\em Pagerank (PR)} is an iterative graph algorithm. At every iteration,
\textit{pagerank\_calculate} function traverses all vertices of a given web graph and
calculates the PageRank of an individual vertex by summing 
weighted
PageRank values of its neighbors.  

All benchmarks output (possibly multi-dimensional) numeric values. To quantify
the accuracy loss in the end result under DPS, we use mean relative error
(i.e., average relative error over all
data points in the output) with
respect to the full-precision outcome. 
The data points in BS are final
stock option prices; in FA, final
positions of cells; in HS, temperature values; in PF,
the final position of the object being tracked;  in PR, the
PageRank values of each vertex. 

As explained in Section~\ref{sec:policy}, in our experiments time advances with each dynamic
function call -- in other words, we use dynamic function calls -- every call to the
aforementioned floating point heavy functions -- as (not
necessarily always homogeneous) units in time.

We experiment with single-threaded
binaries. For FA, PS and PR, the functions subject to DPS distribute
work to threads, hence DPS can expand to parallel execution in a seamless
fashion, by imposing the same precision reduction over all threads. For BS and
HS, on the other hand, the functions subject to DPS can run in parallel, so
different threads may invoke the same function with different precision
simultaneously. We leave further exploration to future work. 

\subsection{Simulation Infrastructure}
We implement the proof-of-concept {\em offline profiler}
compromising statistical fault injection 
and DPS policies using Pin~\cite{pin}, 
as an extension 
to the Pin-based approximate computing framework iACT~\cite{axcPin}. 
During {\em offline profiling}, we inject two types of faults in the mantissa: We set one
mantissa bit (out of 23 for single; 52, for double precision) to 0 ({\em
stuck-at-0}) or 1 ({\em stuck-at-1}) at a time. 
Our tool
instruments all floating point arithmetic and load/store instructions 
for DPS(+). 
\redHL{We compiled the benchmarks using gcc4.9, and disabled SIMD extensions. 
In accordance with the energy model presented in~\cite{BrooksEPI}, we
experimented with a Xeon-Phi like core of 32KB L1
data cache and 512KB L2 data cache.}

\subsection{Energy Model} 
To model energy, we use sum of products of energy per instruction ($EPI$) and
number of instructions ($\#instructions$), over each instruction category (in
each
dynamic function call).
EPI estimates come from measured 
data from~\cite{BrooksEPI}, which 
categorize instructions according to the sources
of operands as {\em RF} (register file), {\em L1}, {\em L2} (level-1 or -2
cache) and (main) {\em memory}.
\cite{BrooksEPI} not only provides robust, measurement based EPI estimates, but
also is 
suitable for our exploration based on the x86 instruction set architecture (due
to the utilization of Pin). In this work, we do not consider vector
optimizations, and 
use scalar
operation EPI values from~\cite{BrooksEPI}, as summarized in
Table~\ref{tab:brooksEPI}.

\begin{table}[htb]
\centering
\caption{EPI values used in this study.}
\label{tab:brooksEPI}
\scalebox{0.93} {
\begin{tabular}{|l|c|}
\hline
\textbf{Instruction Category} ($C$) based on operand source &  {$EPI_C$ (nJs)} \\ \hline
\textbf{RF (register file)}                    & 0.45                        \\ \hline
\textbf{L1}                          & 0.88                        \\ \hline
\textbf{L2}                        & 7.72                        \\ \hline
\textbf{Memory Read (Rd) (with prefetch)}         & 52.14                      \\ \hline
\textbf{Memory Write (Wr) }              & 62.14                       \\ \hline
\end{tabular}
}
\end{table}

To calculate the energy consumption of each dynamic function call, we first
determine the number of instructions in each (operand source based) category,
following the classification in Table~\ref{tab:brooksEPI}. 
Under full precision, the energy consumed by each category in a dynamic function
call becomes $$EPI_C \times \#instructions_C$$ where $C$ represents the
category, i.e., {\em RF}, {\em L1}, {\em L2} or {\em Memory (Rd or Wr)};
$EPI_C$, the EPI estimate of the instructions in category $C$, and
$\#instructions_C$, the number of instructions in the dynamic function call
which fall into category $C$.

Notice that not all of the floating point instructions are subject to precision
reduction under DPS(+). $EPI_C$ values of instructions which keep full precision
directly come from Table~\ref{tab:brooksEPI}~\cite{BrooksEPI}.  For the
instructions subject to omission of mantissa bits of their operands under
DPS(+), on the other hand, EPI changes as a function of the number of (mantissa)
bits omitted -- we will refer to this function as $EPI_{C,o}$. Hence, the energy consumed
by each category $C$ in a dynamic function call becomes $$EPI_C \times
\#instructions_{C,no} + EPI_{C,o} \times
\#instructions_{C,o}$$
$$~~~\texttt{with}~~~\#instructions_{C}=\#instructions_{C,no} + \#instructions_{C,o}$$ 
where
$\#instructions_{C,no}$ ($\#instructions_{C,o}$) represents the number of
instructions in the dynamic function call which fall into category $C$ and where
no (a subset of the) operand mantissa bits are omitted. $EPI_{C,o}$ changes with
the number of omitted bits.

\begin{figure}[h]
  \begin{center}
	\subfloat[Single Precision]{
	  \includegraphics[width=0.5\columnwidth]{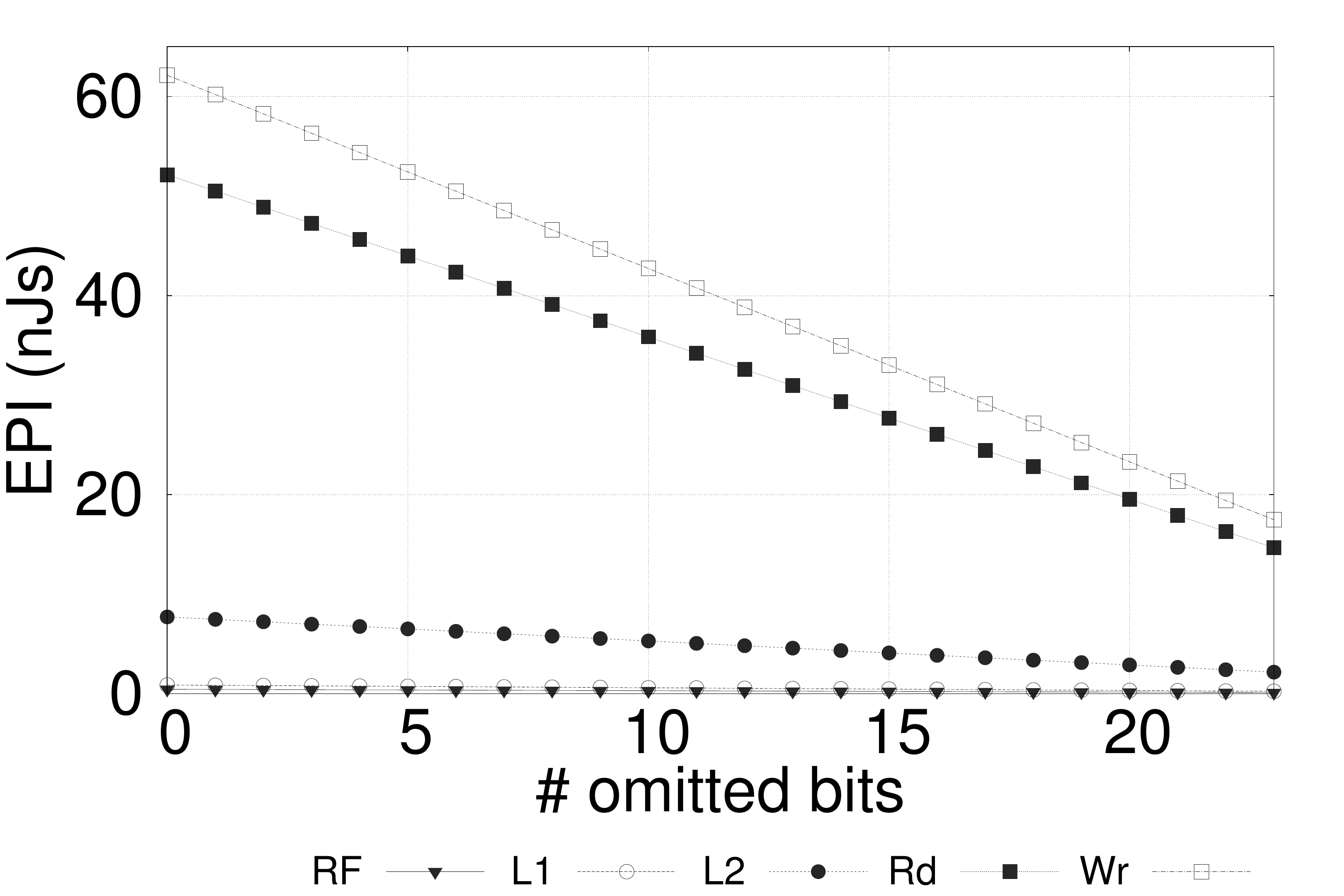}
    }
	\subfloat[Double Precision]{
	  \includegraphics[width=0.5\columnwidth]{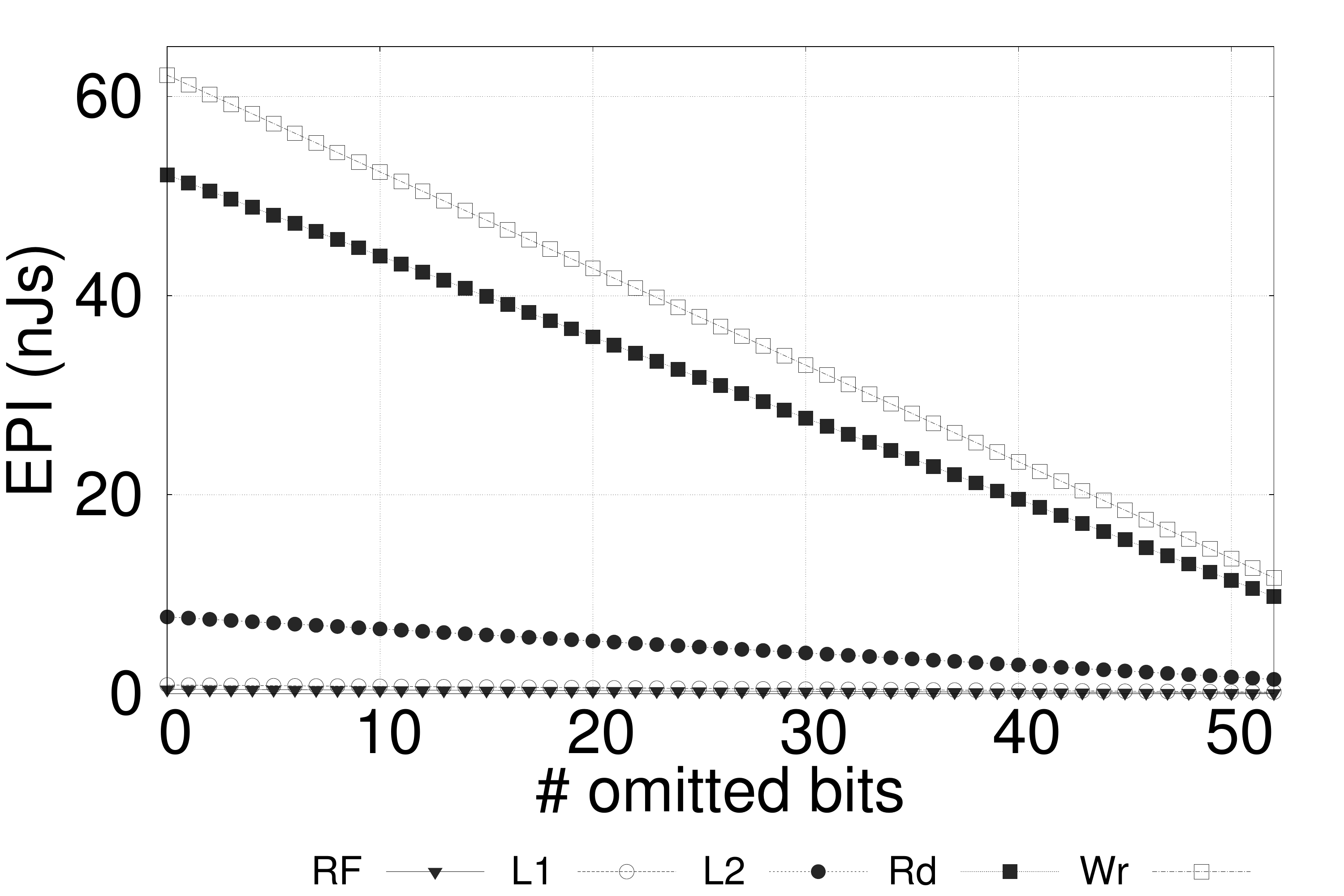}
    }
	\caption{EPI as a function of the number of mantissa bits omitted.
	  \label{fig:scaled_energy_gain}}
  \end{center}
\end{figure}

We deploy the scaling model from~\cite{rutenbar} to model how $EPI_{C,o}$
changes as a function of the number of mantissa bits omitted.
In~\cite{rutenbar}, the authors show that in 
floating point multiplication, which represents one of the most energy hungry
floating point operations, processing 
mantissa bits can easily consume more than $\approx$80\% of the total energy.
\redHL{This study provides a first order analysis of the energy impact of
  reducing mantissa precision, considering energy per operation.
The authors implement a digit-serial multiplier and extract its energy
consumption from SPICE simulation.
The key finding is that energy per operation 
increases linearly with the operand bit-width. This multiplier can also serve 
as a mantissa multiplier.} In this case, the energy consumption increases mostly
linearly with the number of mantissa bits.  In the following, we stick to this
linear model.  
Fig.~\ref{fig:scaled_energy_gain} shows how $EPI_{C,o}$ changes with
number of omitted mantissa bits, for each category $C$, considering single (a) and
double (b) precision.

The linear scaling model from Fig.~\ref{fig:scaled_energy_gain} provides enough
confidence for a limit study to quantify the power efficiency potential for
DPS(+), the goal of this paper.
Accordingly, we do not tie our evaluation to any specific hardware
implementation.
While the picture in Fig.~\ref{fig:scaled_energy_gain} is likely to hold
asymptotically, it will change depending
on whether the underlying hardware
features functional units
of reduced precision or functional units of reconfigurable precision. 

In the following, we will report the cumulative energy savings in the ROI of the benchmark
applications (where the actual computation takes place) under DPS(+), and not
just in
the floating point datapath.  

\begin{figure*}[htp]
  \begin{center}
	\subfloat[BS]{
	  \includegraphics[width=0.35\columnwidth]{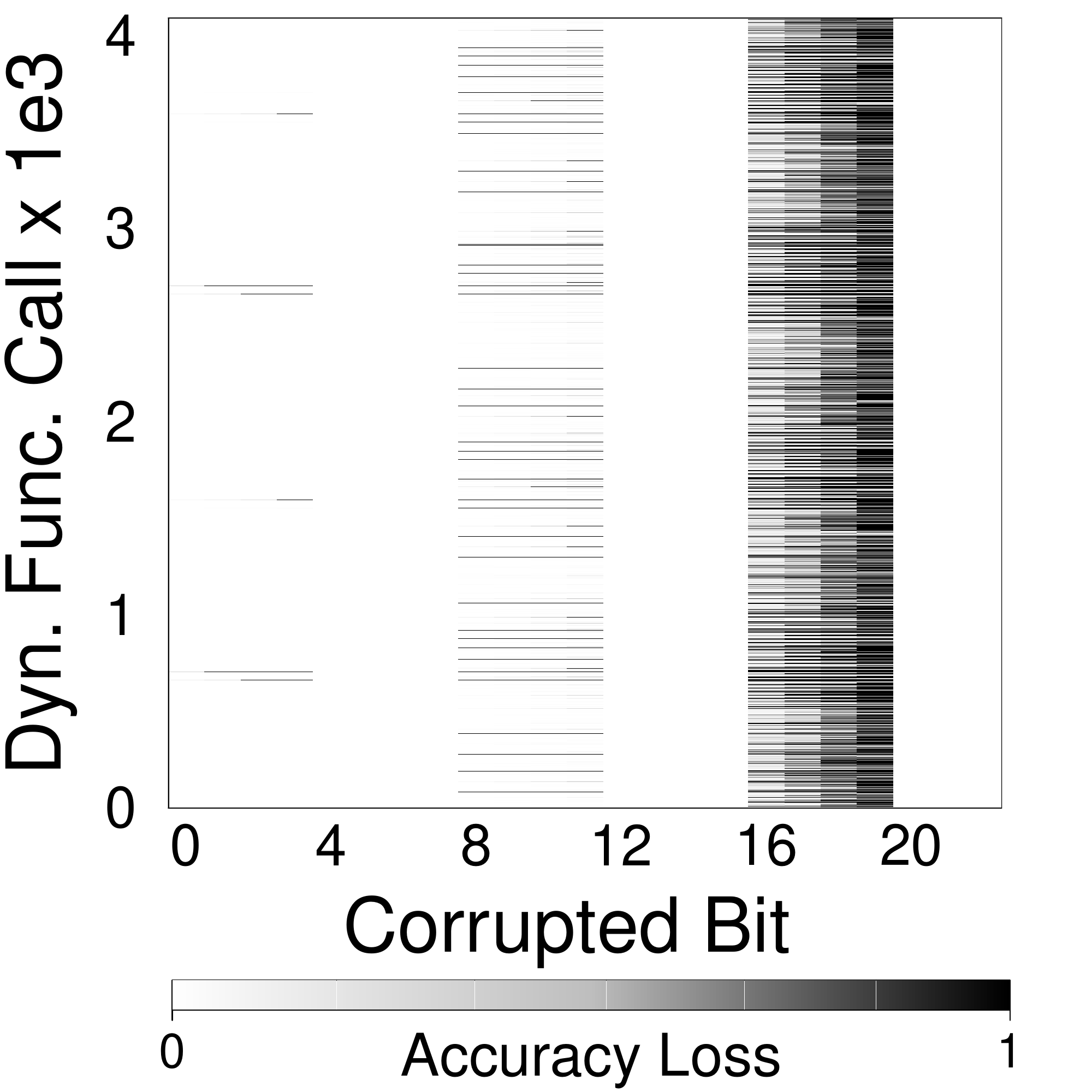}
    }
	\subfloat[FA]{
	  \includegraphics[width=0.35\columnwidth]{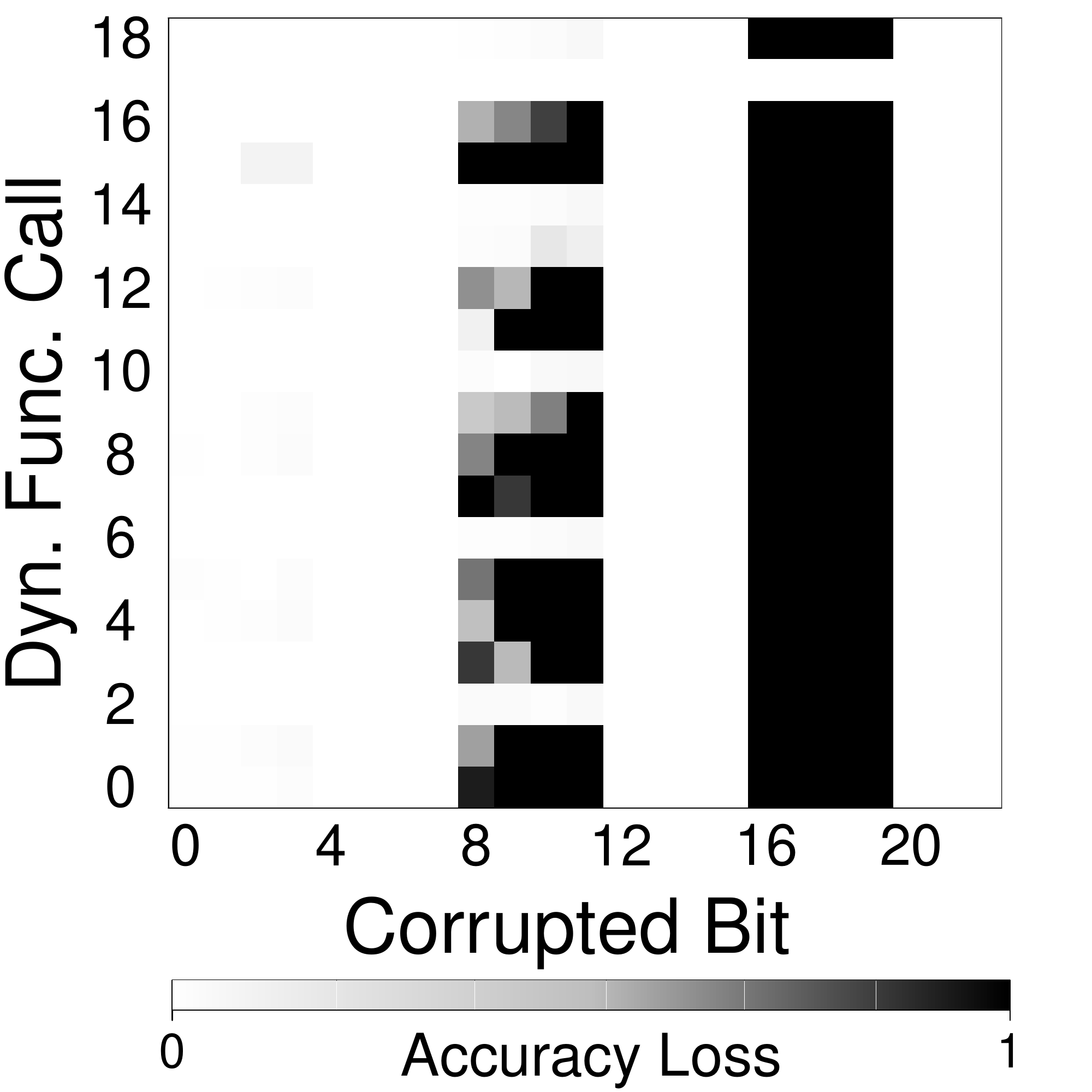}
    }
	\subfloat[HS]{
	  \includegraphics[width=0.35\columnwidth]{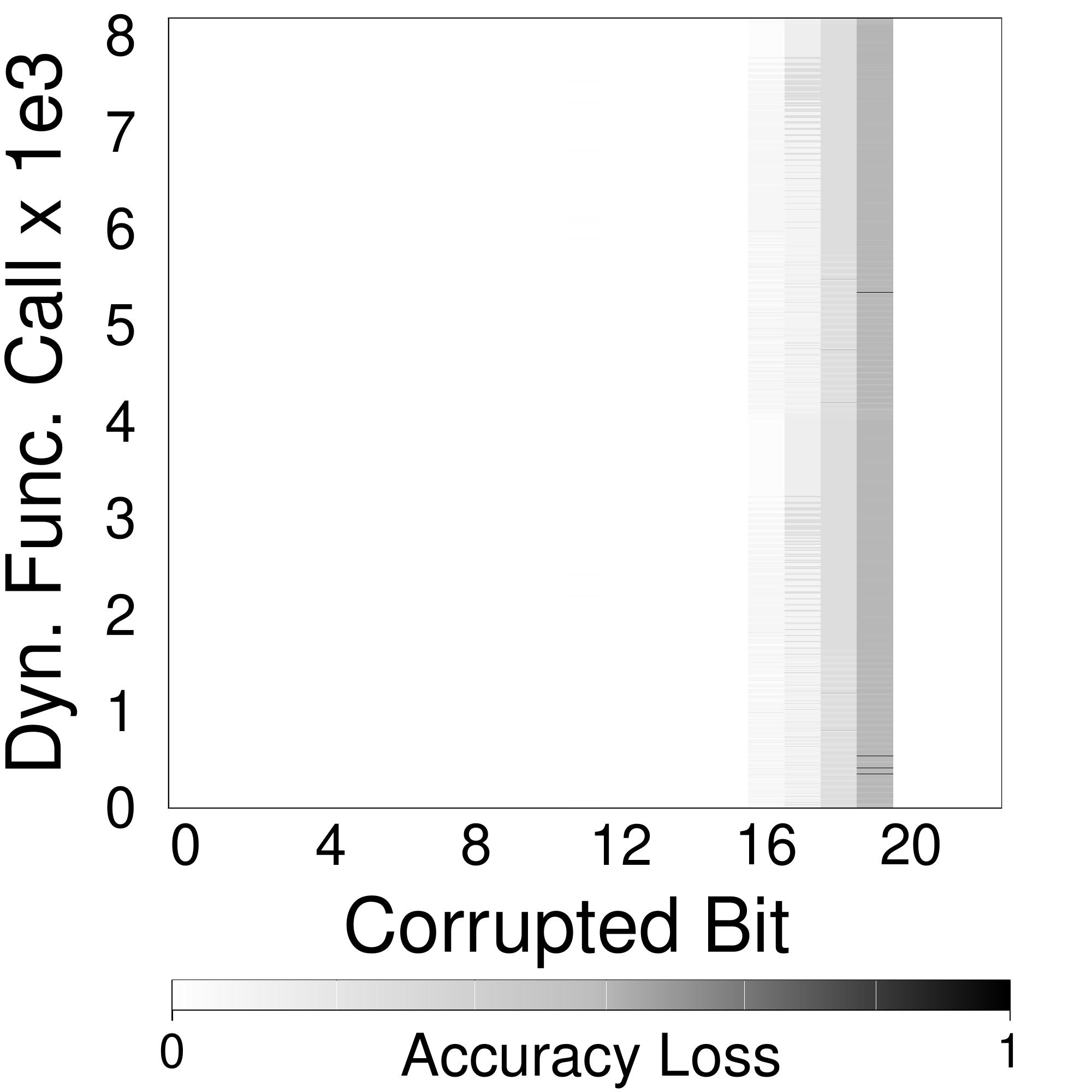}
    }\\
	\subfloat[PR]{
	  \includegraphics[width=0.35\columnwidth]{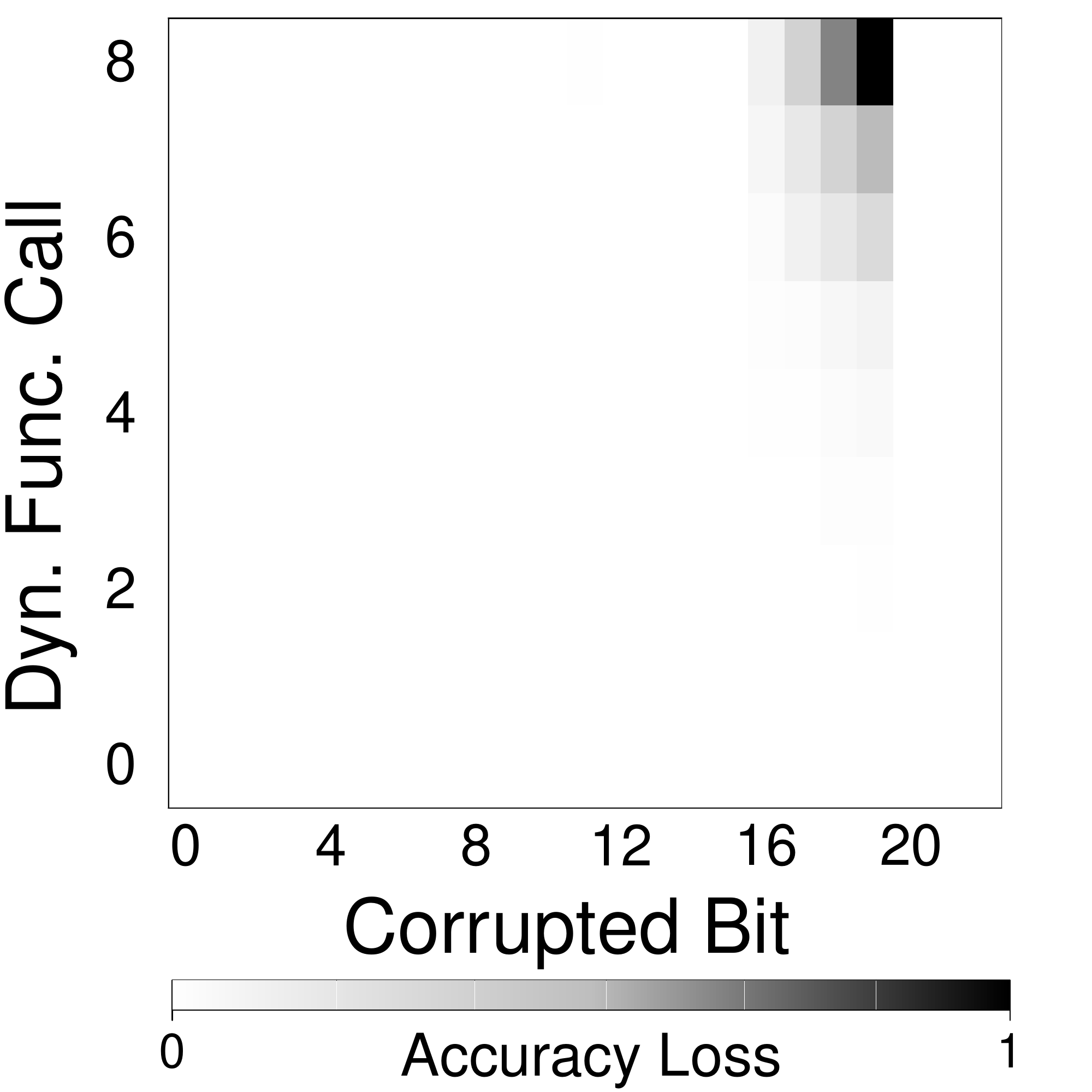}
    }
	\subfloat[PF]{
	  \includegraphics[width=0.35\columnwidth]{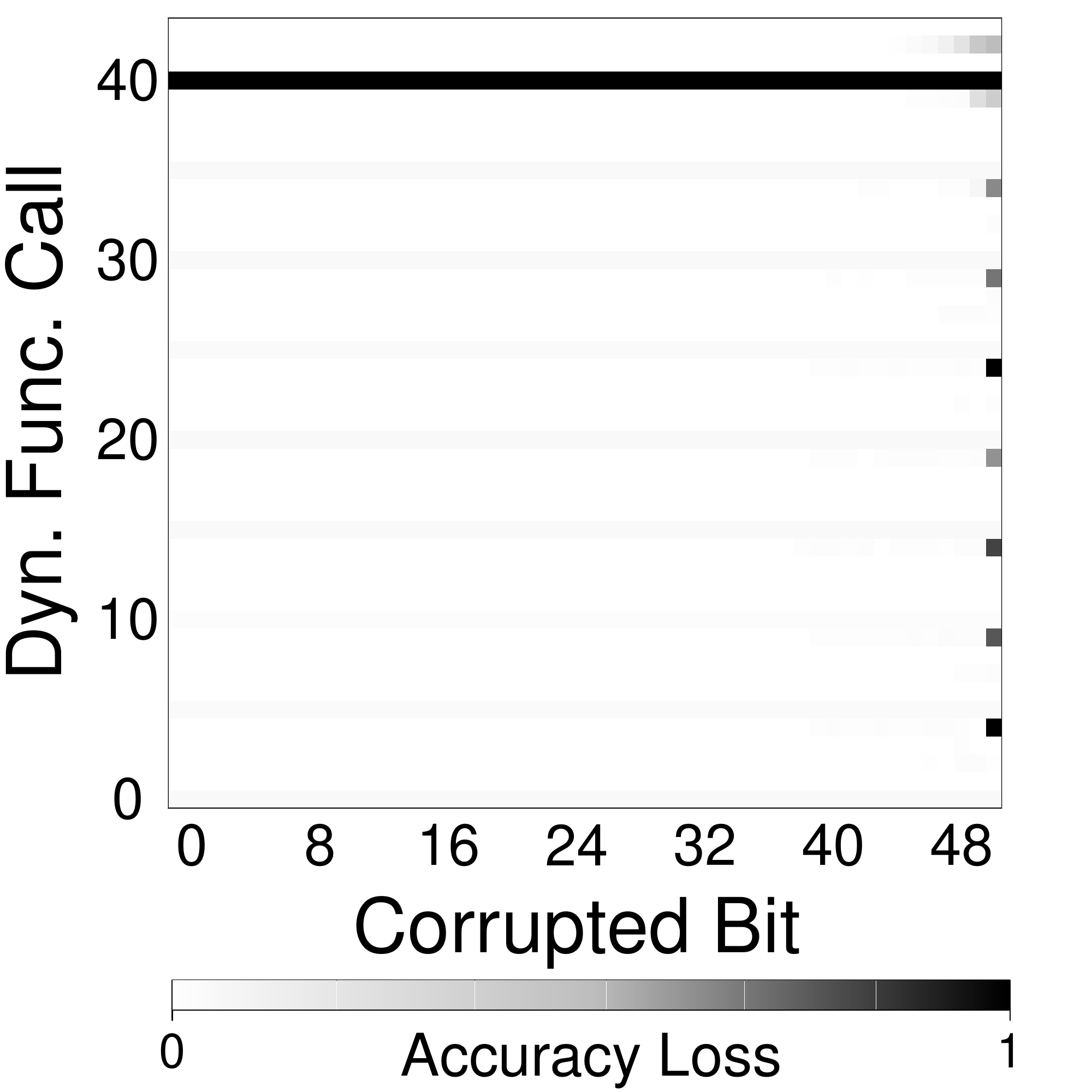}
    }
	\caption{
	  \redHL{Statistical fault injection outcome for temporal noise tolerance. Bit 23
	  on the x-axis demarcates the most significant; bit 0, the least
	significant.}
	  \label{fig:profile}}
  \end{center}
\end{figure*}

\section{Evaluation}
\label{sec:eval}
\subsection{Application Characteristics}
\noindent 
Table~\ref{tab:benchmarks_char} captures the share of  F(loating) P(oint)
operations in the instruction mix and energy consumption of the RMS benchmarks
deployed, over the entire region of interest (ROI).
We observe that the energy share of FP operations in ROI ranges from 12\%
(PF) to 29\% (BS). These numbers provide an upper bound for energy savings under
DPS(+).

\begin{table}[ht]
\centering
\caption{Ratio of FP operations in instruction mix \& energy consumption.}
\label{tab:benchmarks_char}
\scalebox{0.95} {
\begin{tabular}{|l|c|c|}
\hline
\textit{\textbf{Benchmark}} & \textbf{FP ops. in ROI} & \textbf{FP energy in ROI} \\ \hline
\textbf{BS}            &   0.25 &    0.29                                \\ \hline
\textbf{FA}           &    0.15     & 0.13 \\ \hline
\textbf{HS}            & 0.12 &  0.17    \\ \hline
\textbf{PF}         &         0.10            &   0.12   \\ \hline
\textbf{PR}           &  0.05   &  0.13     \\ \hline
\end{tabular}
}
\end{table}

Fine grain temporal changes in the noise tolerance of RMS applications motivates
DPS(+). Fig.~\ref{fig:profile} verifies this insight, where we plot the outcome
of the fault injection experiments to measure the sensitivity of each FP-heavy
dynamic function call to noise.
The x-axis depicts which mantissa bit we corrupt.
The
left y-axis shows the dynamic function calls, in the order of execution (as
identified during profiling). The right y-axis captures the (relative) accuracy
loss in the end result as induced by the corrupted bit in gray scale (black indicates a
totally inaccurate result, i.e., a relative accuracy loss of 1; white, no
accuracy loss).
As expected, we observe more darker regions (less noise tolerance) as we move
right on the x-axis -- as
we corrupt higher order (more significant) mantissa bits.

\begin{figure}[htp]
  \begin{center}
	\subfloat[FA]{
	  \includegraphics[width=0.4\columnwidth]{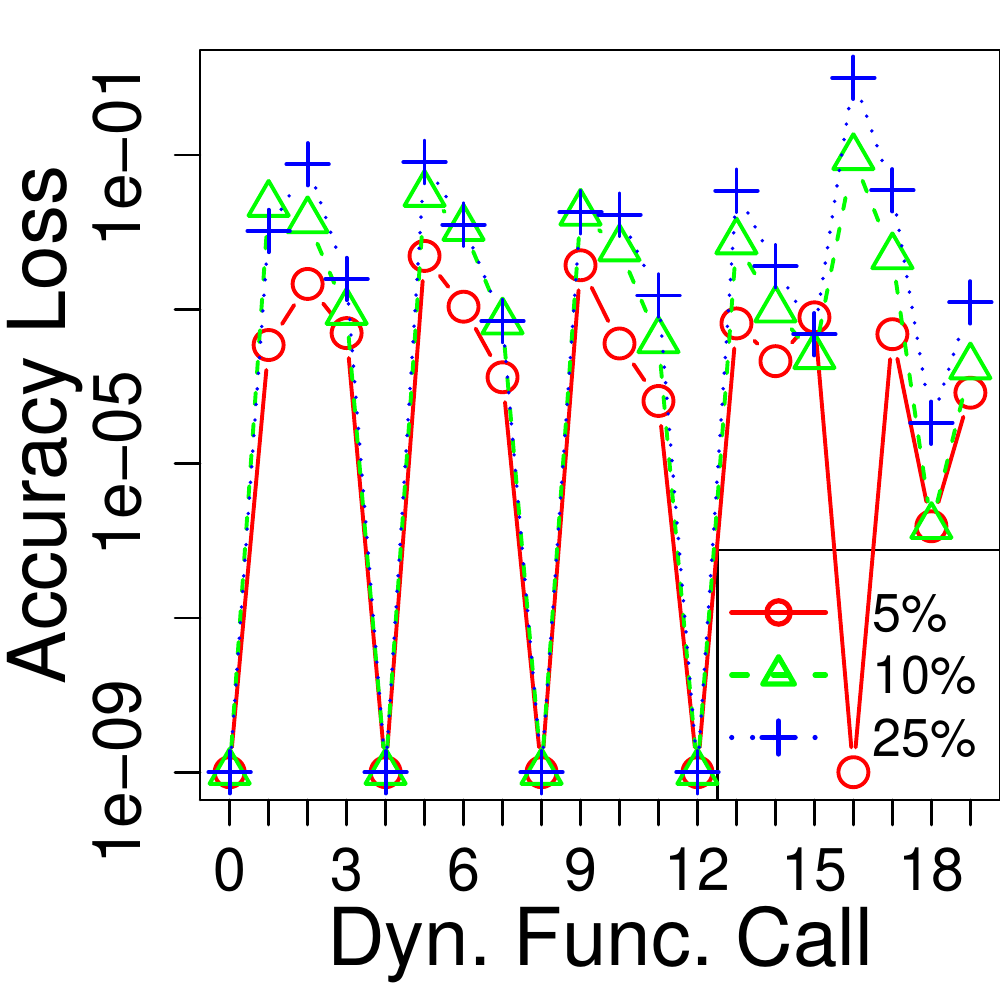}
    }
	\hspace{.5cm}
	\subfloat[PR]{
	  \includegraphics[width=0.4\columnwidth]{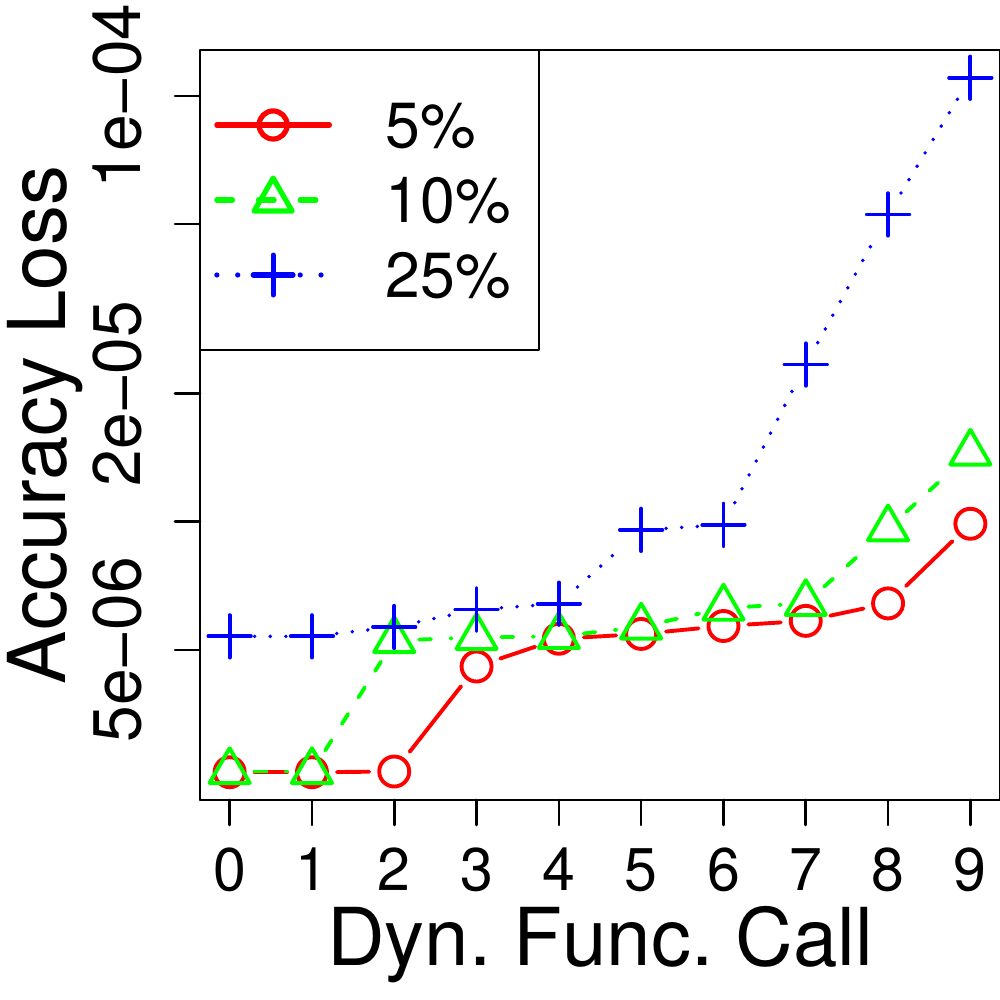}
    }
\caption{Evolution of noise tolerance with time.
	  \label{fig:motiv}}
  \end{center}
\end{figure} 

We observe that the
accuracy loss (right y-axis) indeed changes with time (i.e., with dynamic function
calls as depicted on the left y-axis) for all applications, least pronounced for HS in Fig.~\ref{fig:profile}(c).
For the rest of the applications, the accuracy loss (as a proxy of noise
tolerance) shows recurring patterns over time:
For example, FA (b) periodically enters a relatively more noise tolerant phase
(as characterized by \emph{AdvanceParticle} function);
PF (e), a relatively less noise tolerant phase (as characterized by
\emph{apply\_motion\_model} function).
PR (d), on the other hand, exhibits less noise tolerance as we move up on the
left y-axis -- in
later stages of execution. This is because, relying on iterative refinement, PR
has less opportunities to recover from
noise in later stages of execution.   
Careful inspection reveals barely any  accuracy loss, even due to the corruption
of more significant mantissa bits,
until sixth iteration. 
After sixth iteration, however, we start to observe sizable loss in accuracy.

Most of the time, differences in the noise tolerance of each dynamic call to the
very same static function stem from differences in the function inputs across
calls. 
For instance, BS has a single computational kernel. Each (dynamic) call to
this kernel processes different inputs.  Fig.~\ref{fig:profile}(a) reveals the
differences in accuracy loss among these calls due to inputs along the time (left y-) axis.
Input data values also have an impact. When working with smaller values,
corruptions in least significant bits are more likely to induce higher accuracy
loss in the end result. This applies to the fluctuations in accuracy loss across
the left y-axis due to
the corruption after least significant 8 bits in (a). 
All of these results point to opportunities for DPS.

\begin{figure}[h]
  \begin{center}
	\subfloat[Accuracy Loss]{
	  \includegraphics[width=0.45\columnwidth]{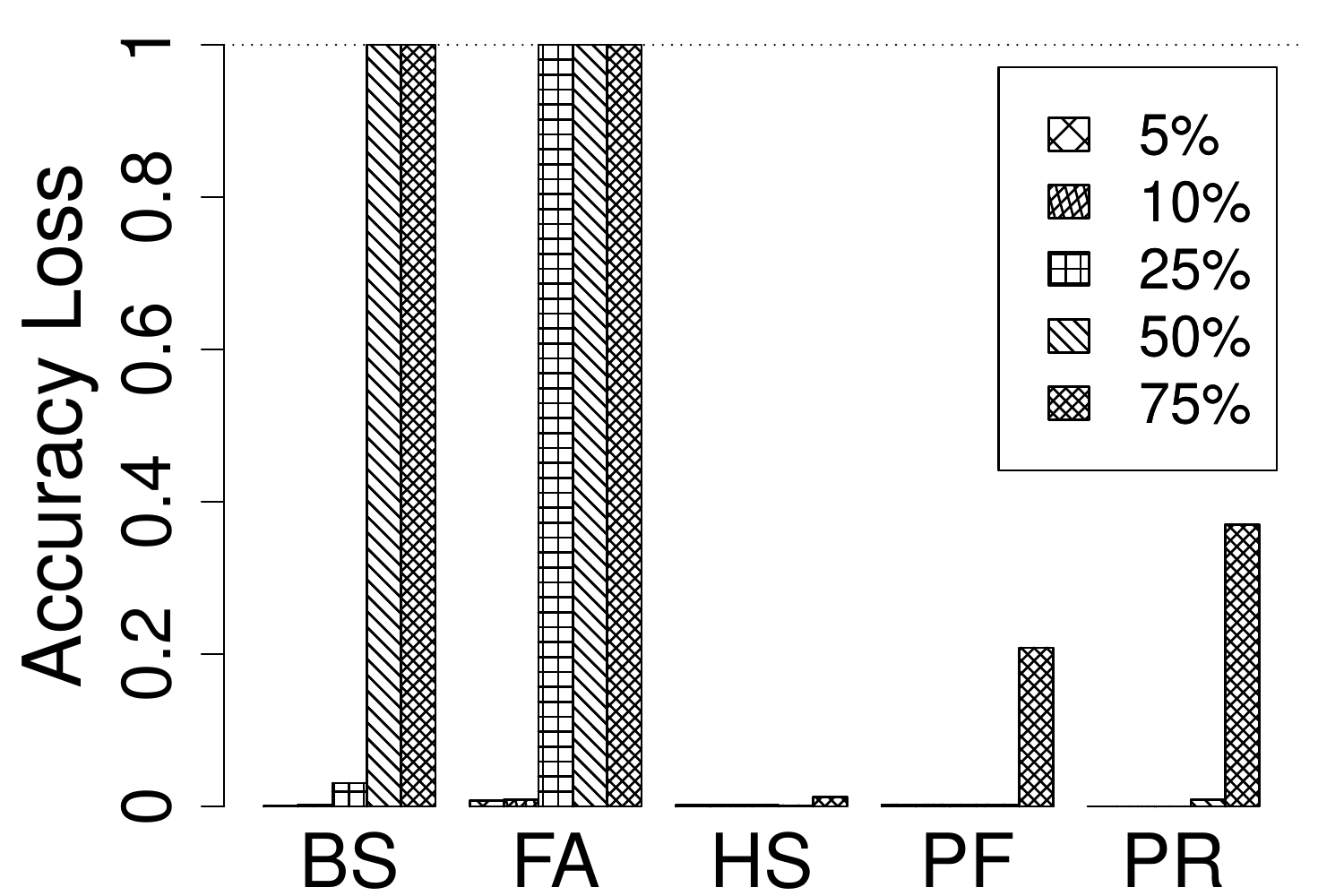}
    }
	\hspace{0.5cm}
	\subfloat[Energy]{
	  \includegraphics[width=0.45\columnwidth]{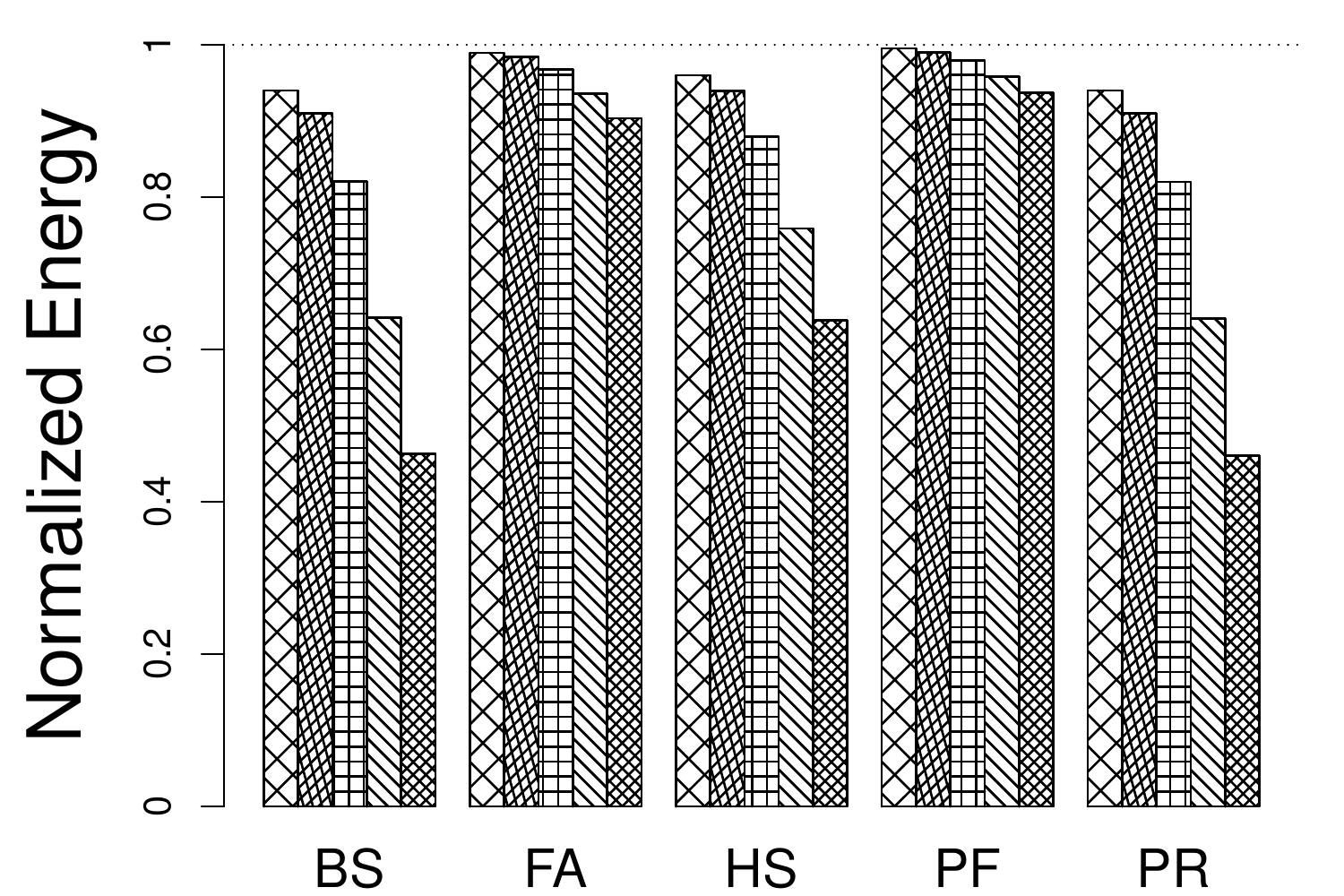}
    }
	\caption{Impact of SPS on accuracy loss (a) and energy (b).
	  \label{fig:static}}
  \end{center}
\end{figure}

Fig.~\ref{fig:motiv} provides the microscopic view for two of the applications
from Fig.~\ref{fig:profile}: FA (a) and PR (b).
The x-axis corresponds to time (the left y-axis of Fig.~\ref{fig:profile});
the y-axis, to relative accuracy loss (the right y-axis of
Fig.~\ref{fig:profile}).
In this case, we omit multiple consecutive mantissa bits; specifically,  (ceiling of) 5\%,
10\%, and 25\% of mantissa bits, starting from the least significant.
In line with our findings from Fig.~\ref{fig:profile}, we identify how the noise
tolerance of FA (a) fluctuates; and of PF (b), decreases over time.

\subsection{Static Precision Scaling (SPS)}
\label{sec:sps}
\noindent We next explore how the accuracy loss and energy consumption look like if
we impose a fixed degree of precision reduction, statically, throughout the
entire execution. In the following, we will refer to this policy as {\em Static
Precision Scaling (SPS)}.
We will use the outcome under SPS 
as a baseline for comparison. 
As SPS does not differentiate between noise tolerance of dynamic calls, the
least noise tolerant function is likely to determine 
the final accuracy loss.
Under SPS, we omitted (ceiling of) 5\%-75\% of mantissa bits. 
Fig.~\ref{fig:static} captures the outcome.
We observe that SPS can render sizable energy savings (b), particularly as we
omit more than 10\% of the bits. However, for BS and FA the energy savings are
accompanied by the excessive accuracy loss as revealed in (a).
According to Fig.~\ref{fig:profile}, FA can tolerate 
corruption in higher order bits of mantissa, but SPS cannot unlock this
opportunity.
Under SPS, FA renders unacceptable accuracy loss if we omit 25\% of the bits (6
bits). In the next section, we will show that FA can tolerate omission of up to
15 bits under DPS (+) (Fig.~\ref{fig:pol_fa}). Similarly, under SPS, BS cannot
tolerate the omission of 
50\% of its mantissa bits (12 bits), where according to Fig.~\ref{fig:profile},
many of its dynamic calls
may tolerate corruption at higher order bits (bit 18, e.g.).
For the rest of the applications, SPS performs arguably well. Still, DPS(+) can
unlock more opportunities for power efficiency: 
As a
specific example, PR under SPS with 75\% of mantissa bits (18 bits) omitted
results in 0.37$\times$ accuracy loss. According to Fig.~\ref{fig:profile}, 
PR
can temporally tolerate the omission of higher order bits. In the next Section we will show
how DPS(+) can unlock this opportunity by omitting 90\% of the mantissa bits on
average to render an accuracy loss of 
0.13$\times$.

\subsection{Dynamic Precision Scaling (DPS)}
\begin{figure}[h]
  \begin{center}
	\subfloat[DPS]{
	  \includegraphics[width=0.4\columnwidth]{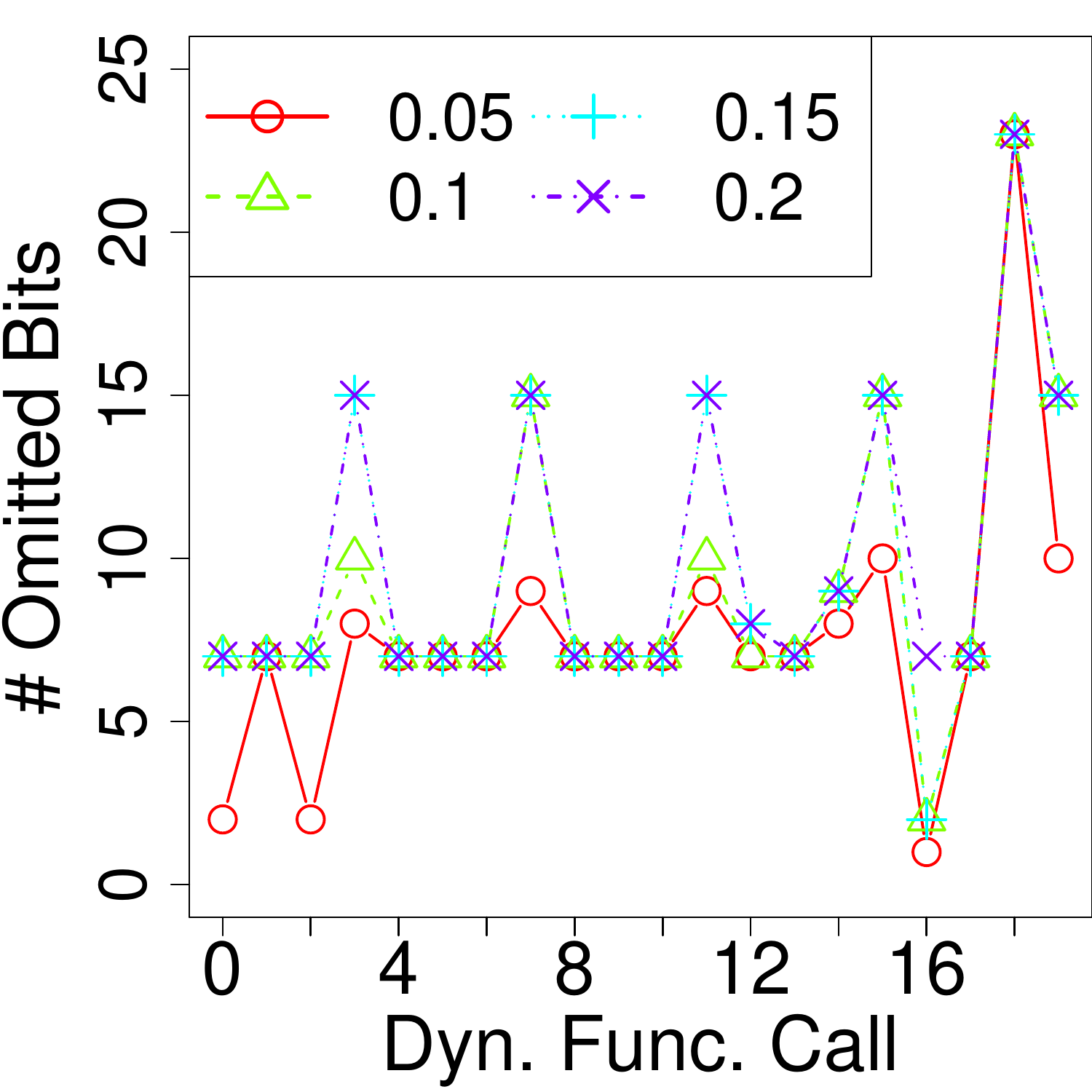}
    }
	\hspace{.5cm}
	\subfloat[DPS+]{
	  \includegraphics[width=0.4\columnwidth]{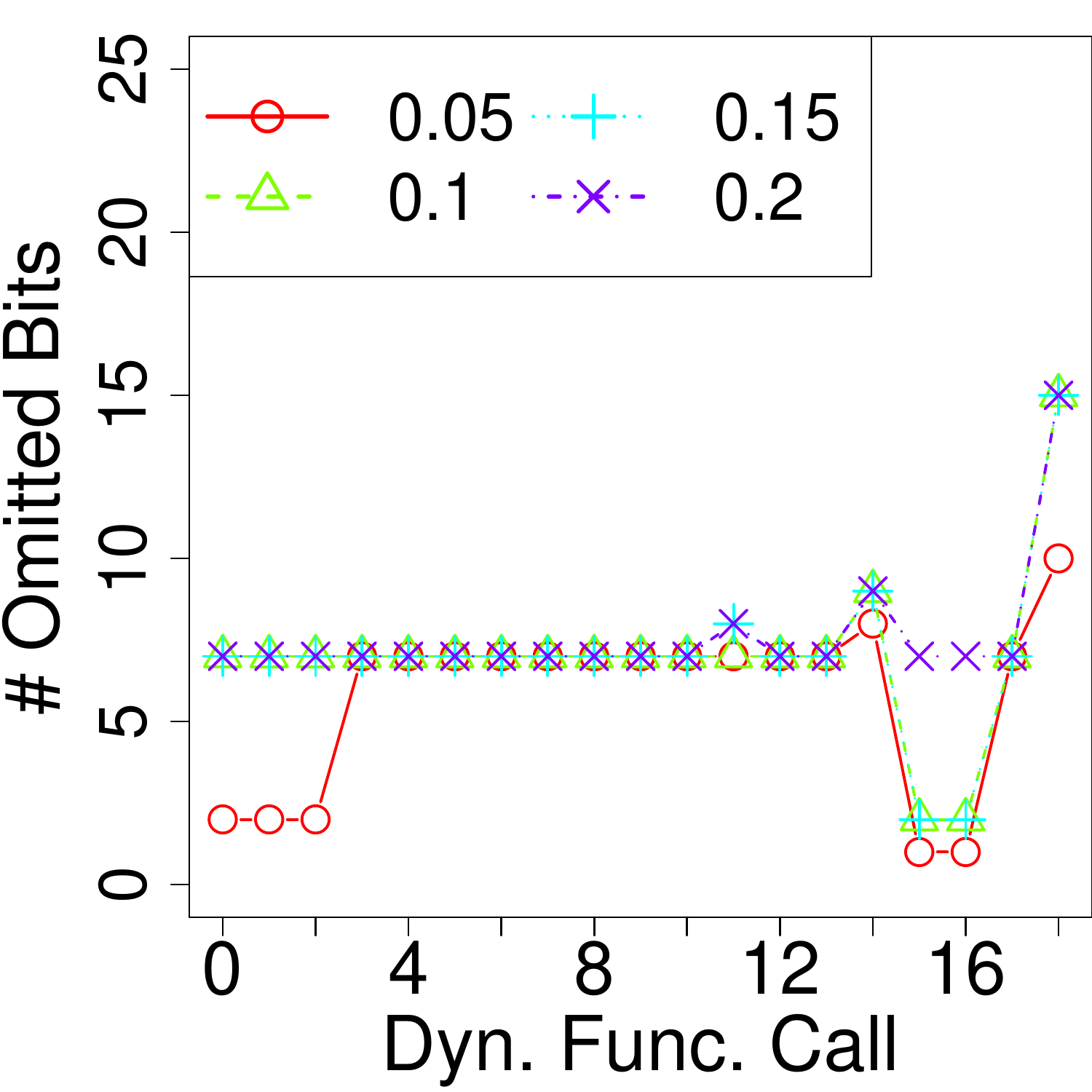}
    }
	  \caption{\#omitted bits under DPS(+)  for FA.
	  \label{fig:pol_fa}}
  \end{center}
\end{figure}

\begin{figure}[h]
  \begin{center}
	\subfloat[DPS]{
	  \includegraphics[width=0.4\columnwidth]{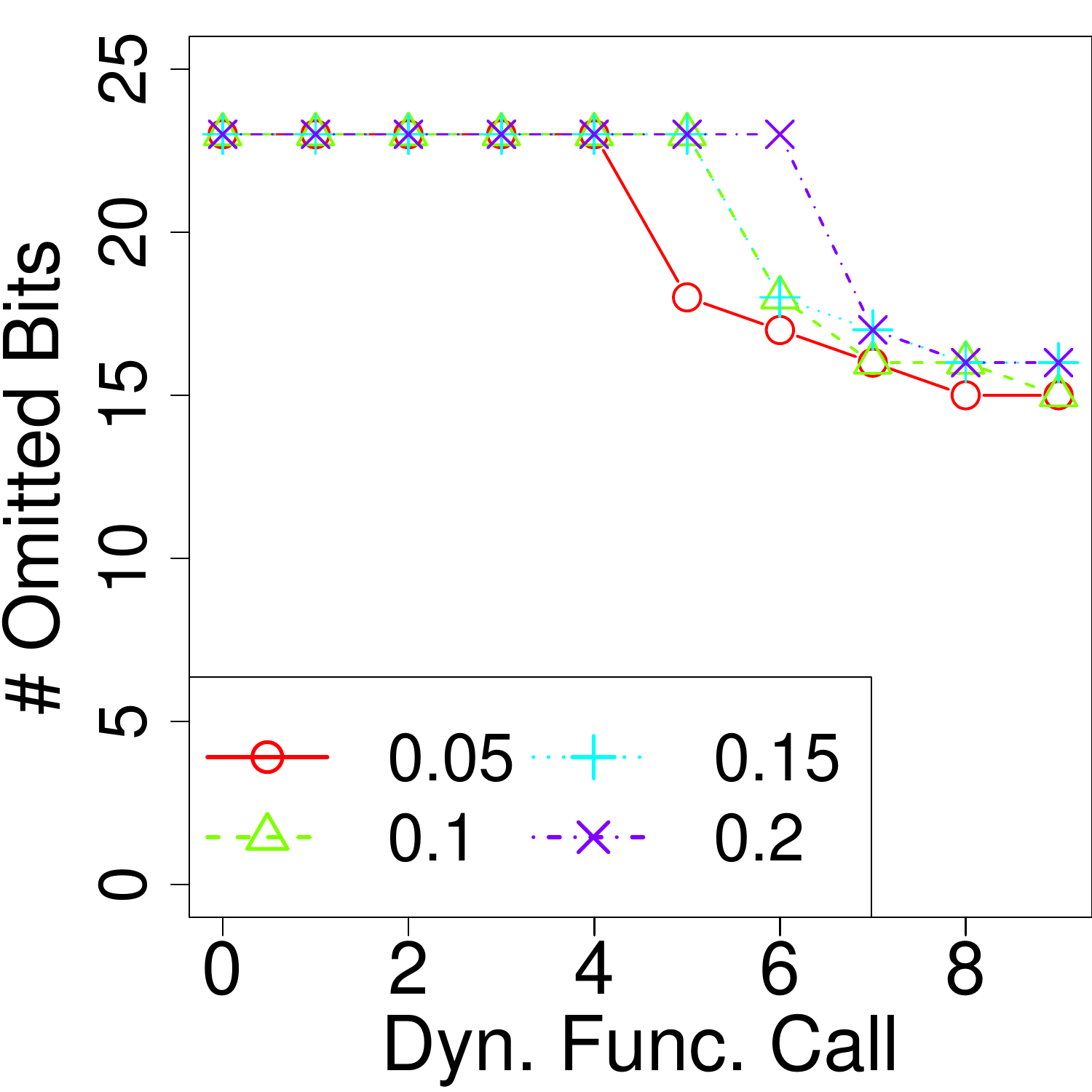}
    }
	\hspace{0.5cm}
	\subfloat[DPS+]{
	  \includegraphics[width=0.4\columnwidth]{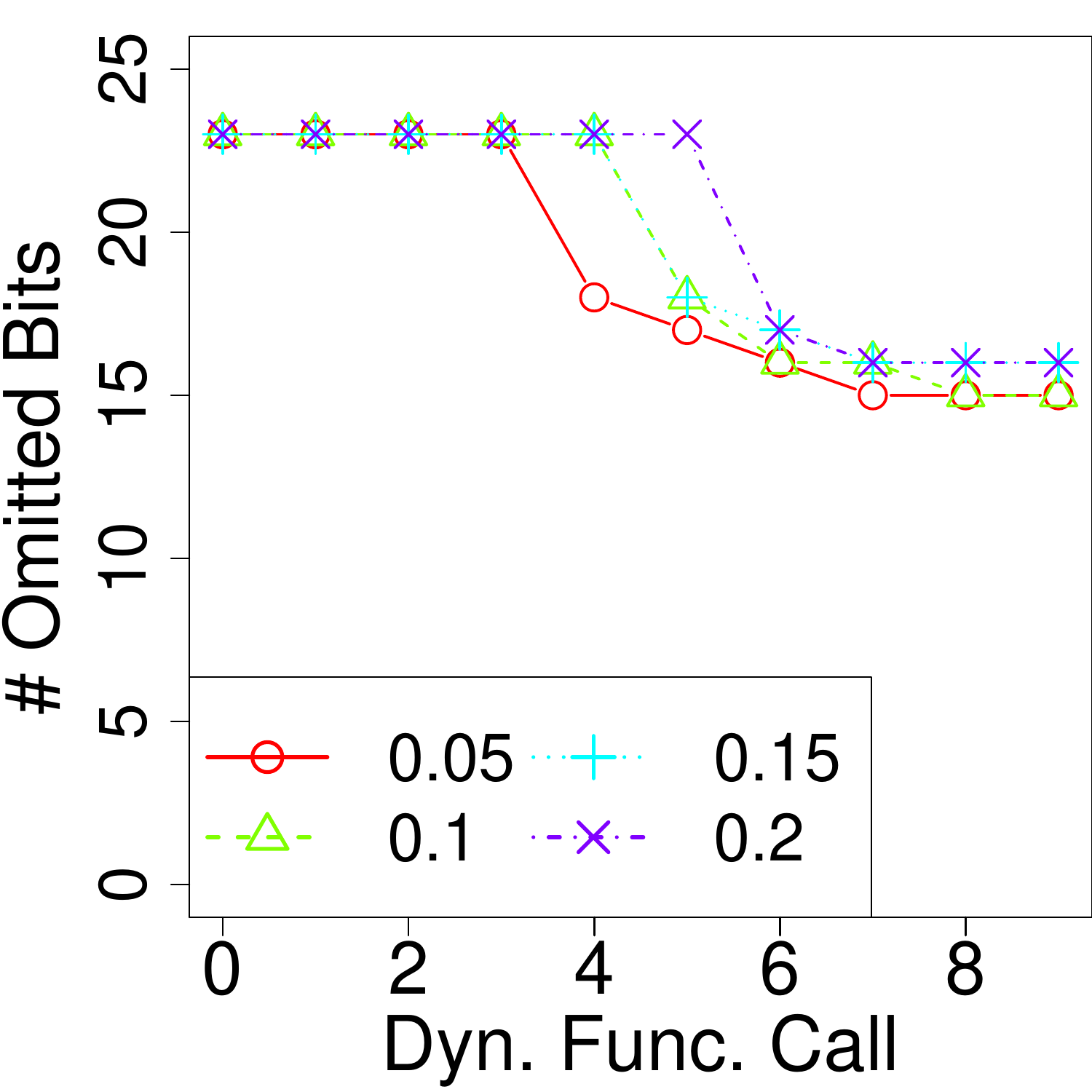}
    }
	\caption{\#omitted mantissa bits under DPS(+) for PR.
	  \label{fig:pol_pr}}
  \end{center}
\end{figure}

\begin{figure}[h]
  \begin{center}
	\subfloat[DPS]{
	  \includegraphics[width=0.4\columnwidth]{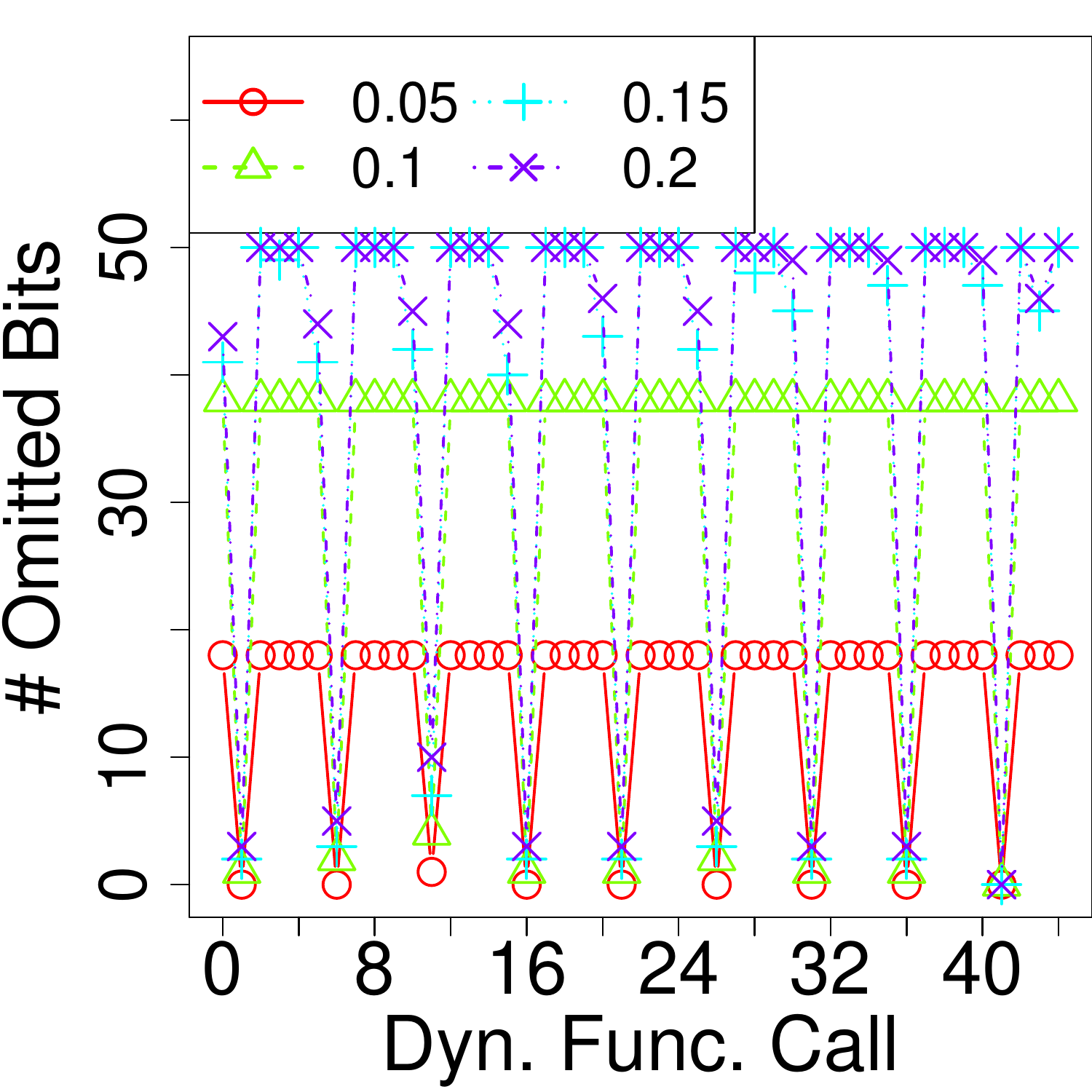}
    }
	\hspace{0.5cm}
	\subfloat[DPS+]{
	  \includegraphics[width=0.4\columnwidth]{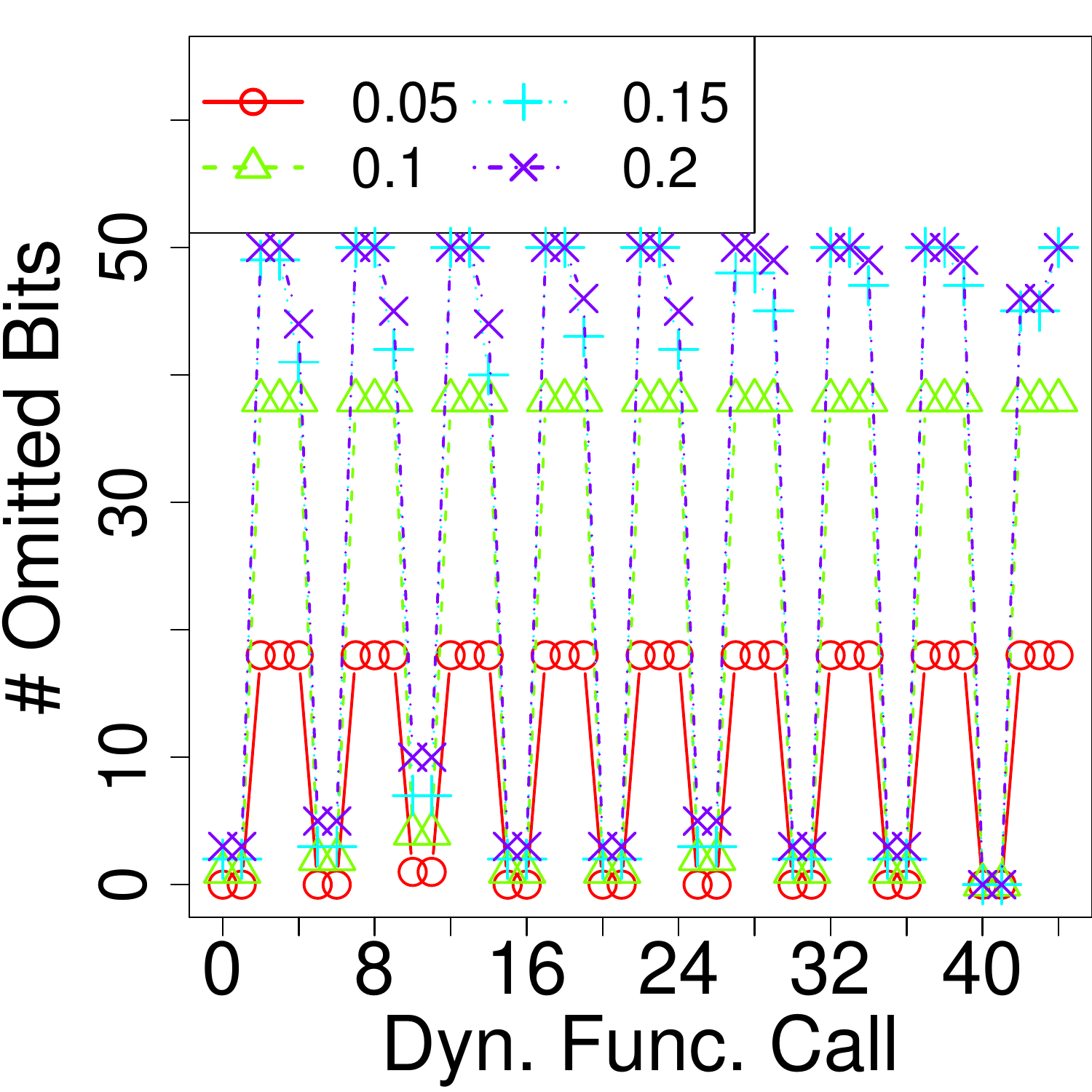}
    }
	\caption{\#omitted bits under DPS(+) for PF.
	  \label{fig:pol_pf}}
  \end{center}
\end{figure}

We next evaluate the effectiveness of DPS(+).
We invoke the two DPS algorithms with different values of {\em targetAccLoss}
(Section~\ref{sec:policy}),
the maximum accuracy loss in the end result the application can tolerate, and
feed the resulting {\em \#omittedBits} to the {\em runtime monitor}. Recall that 
{\em \#omittedBits}
gives the total number of
(consecutive) mantissa bits (starting from the least significant) we can omit
on a per dynamic call basis, while the corresponding accuracy loss in end result remains lower than {\em
targetAccLoss}.
In the following, we report the outcome for {\em targetAccLoss} values
between 0.05 and 0.2 
at increments of 0.05.
\noindent Fig.s~\ref{fig:pol_fa},~\ref{fig:pol_pr}, and~\ref{fig:pol_pf} depict the
number of omitted bits for each dynamic function call, as a function of {\em
targetAccLoss}, for FA, PR, and PF, respectively; under DPS in (a), and DPS+ in
(b). 
\ignore{
\redHL{In addition to DPS and DPS(+), we also evaluate DPS\_min as a baseline.
DPS\_min works as follows: (1) we run DPS heuristic to find omitted bits for
each dynamic function call. (2) then, we find the minimum number of bits omitted
for each static function by using its dynamic function calls and number of
omitted bits pairs.}
}

The pattern under DPS closely tracks our findings in
Fig.~\ref{fig:profile}, as DPS considers each dynamic call in isolation. This
observation also holds for BS and HS, not shown as the corresponding figures
were barely readable due to very fine grain temporal fluctuations.
Recall that the x-axis captures each dynamic function call in the order of
execution, and hence represents a proxy for time.
Fig.s~\ref{fig:pol_fa},~\ref{fig:pol_pr}, and~\ref{fig:pol_pf},
respectively, for FA, PR, and PF, show how the 
number of omitted bits changes over time to track the temporal changes in the
noise tolerance of the applications (more noise-tolerant phases being able to
accommodate a higher number of omitted mantissa bits).

We next examine the corresponding accuracy loss in the end result of the
applications in Table~\ref{tab:final_error} for DPS, 
in Table~\ref{tab:final_error_alg2} for DPS+, and in Table~\ref{tab:final_error_minDPS} for DPS\_min. 
We observe that (i) as expected, a higher {\em
targetAccLoss} value renders monotonically higher
accuracy loss for all applications; (ii) the accuracy loss under DPS eventually overshoots {\em
targetAccLoss} (highlighted in bold); (iii) by dependency tracking, DPS+ manages to
eliminate most of the overshoots and to bound the final accuracy
loss
opportunistically. 

\begin{figure}[h]
  \begin{center}
	  \includegraphics[width=0.5\columnwidth]{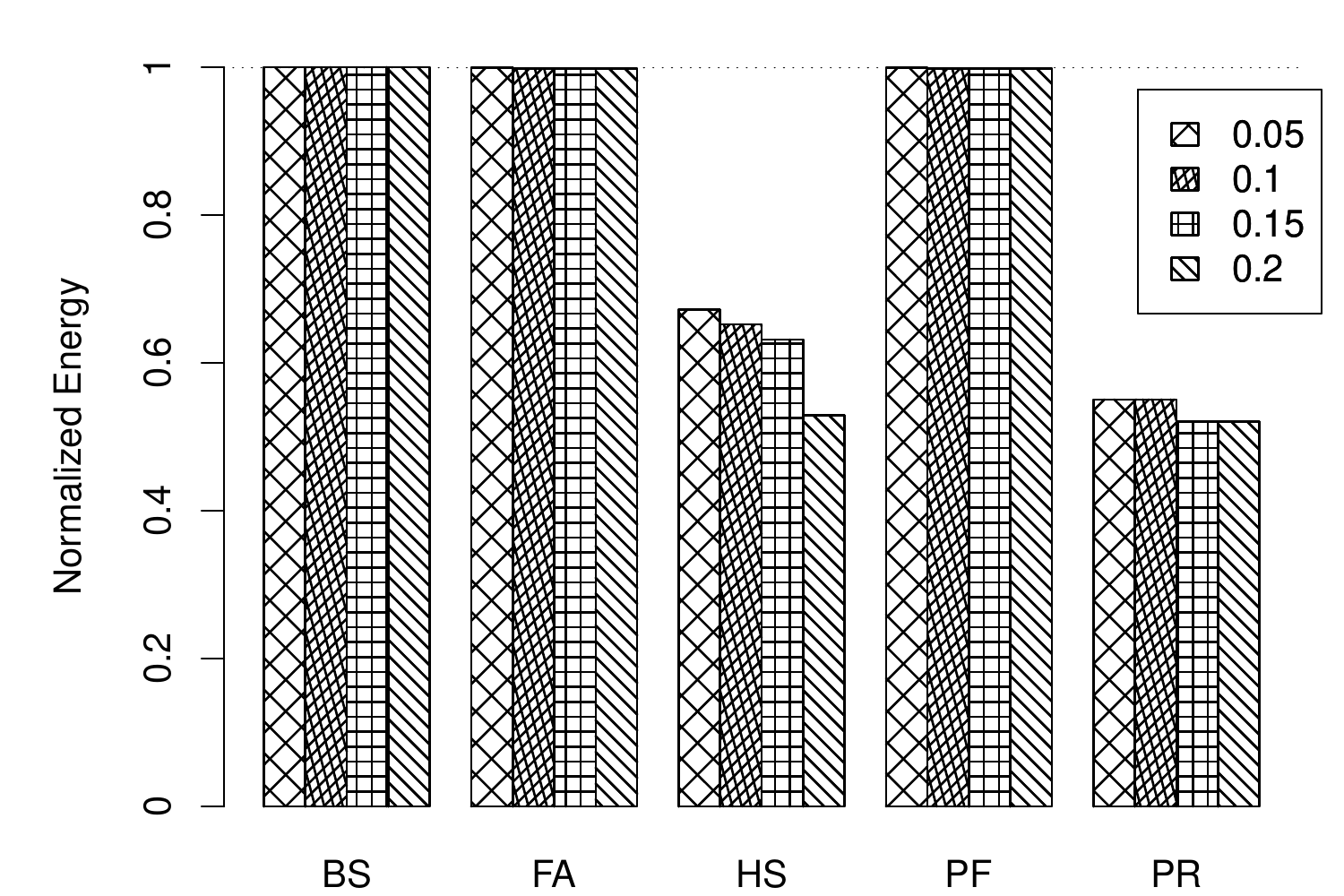}
	\caption{\redHL{Energy consumption under SPS+ for different {\em targetAccLoss} values.}
	  \label{fig:energyMinDPS}}
  \end{center}
\end{figure}

\redHL{At this point, we introduce one more baseline for comparison, SPS+, which can eliminate
such overshoots. SPS+
works as follows: During profiling, (i) we first run DPS heuristic to find the number of bits to be
omitted for
each dynamic function call; (ii) we then find the minimum {\em \#omittedBits}
for each static function by using the profile from (i) over all of its dynamic
function calls. After profiling, we impose this number of bits on all dynamic calls throughout
execution.
As
Table~\ref{tab:final_error_minDPS} reveals, SPS+ renders a lower accuracy loss.
However, as SPS+ cannot exploit temporal changes in algorithmic noise tolerance,
it leaves potential savings in energy untapped.
For example, BS features a single target function, some dynamic instances of
which cannot tolerate approximation (i.e., where {\em \#omittedBits} becomes
zero). SPS+ in this case imposes this {\em \#omittedBits} as the minimum over
all dynamic calls, and hence, excludes any
approximation. As a result, SPS+ cannot deliver any energy savings for BS as
opposed to DPS(+), as we will see shortly in Figure~\ref{fig:energyMinDPS}.
}

\begin{table}[ht]
\centering
\caption{Accuracy loss under DPS for different {\em targetAccLoss} values}
\label{tab:final_error}
\begin{tabular}{|l|l|l|l|l|l|l|l|l|}
\hline
            &  \textbf{0.05}& \textbf{0.1} 	& \textbf{0.15} 	& \textbf{0.2} \\ \hline
\textbf{PF} &  0.0025  		& 0.0028 		& 0.0347 			& \redHL{\bf 0.7759} \\ \hline
\textbf{HS} &  0.0036   	& 0.01233 		& \redHL{\bf 0.2140}  			&
\redHL{\bf 0.2140} \\ \hline
\textbf{PR} &  \redHL{\bf 0.0849}  		& \redHL{\bf 0.1416}		& \redHL{\bf 0.1714}
& \redHL{\bf 0.2582} \\ \hline
\textbf{FA} &  5.63E-5     	& 8.25E-5    	& 8.32E-5     		& 0.0823 \\ \hline
\textbf{BS} &  0.02907  	& 0.0486 		& 0.0688  			& 0.0769 \\ \hline
\end{tabular}
\end{table}

\begin{table}[ht]
\centering
\caption{Accuracy loss under DPS+ for different {\em targetAccLoss} values }
\label{tab:final_error_alg2}
\begin{tabular}{|l|l|l|l|l|l|l|l|l|}
\hline
            & \textbf{0.05} & \textbf{0.1} 	& \textbf{0.15} & \textbf{0.2} 	\\ \hline
\textbf{PF} & 0.0025  		& 0.0025 		& 0.0111  		& 0.0111 		\\ \hline
\textbf{HS} & 0.0035   		& 0.0123	 	& \redHL{\bf 0.2120}	  	&
\redHL{\bf 0.2140} 		\\ \hline
\textbf{PR} & \redHL{\bf 0.0758}  		& 0.0848		& 0.1428  		& 0.1646 		\\ \hline
\textbf{FA} & 9.66E-6     	& 2.64E-5    	& 2.64E-5     	& 0.0822	  	\\ \hline
\textbf{BS} & 0.0234  		& 0.0419 		& 0.0605  		& 0.0682 		\\ \hline
\end{tabular}
\end{table}

\begin{table}[ht]
\centering
\caption{\redHL{Accuracy loss under SPS+ for different {\em targetAccLoss} values} }
\label{tab:final_error_minDPS}
\begin{tabular}{|l|l|l|l|l|}
\hline
            & \textbf{0.05}  & \textbf{0.1}   & \textbf{0.15} & \textbf{0.2}  \\ \hline
\textbf{PF} & 0.0025 & 0.0027 & 0.0110 & 0.0130 \\ \hline
\textbf{HS} & 0.0032 & 0.0066 & 0.0123 & \redHL{\bf 0.2140} \\ \hline
\textbf{PR} & \redHL{\bf 0.0615} & 0.0615  &0.1308   & 0.1308   \\ \hline
\textbf{FA} & 8.84E-6       & 8.86E-6       & 3.30E-5      & 0.0822 \\ \hline
\textbf{BS} & 0              & 0              & 0             & 0             \\ \hline
\end{tabular}
\end{table}

PF and PR feature data-dependent
consecutive dynamic function calls. Since the output of the preceding function call 
acts as the input
to the next function call, omitting more bits in the preceding call (than the
following call) tends to result in higher, and possibly excessive
accuracy loss. DPS+ handles this case by matching the precision reduction of
consecutive calls. Mainly applications featuring this or similar type of data
dependency can benefit from DPS+.
As a result, under DPS+, the accuracy loss of PF and PR significantly decreases.
For example, at {\em targetAccLoss}=0.05,
Fig.~\ref{fig:pol_pr}(b) shows how DPS+ matches the number of
omitted bits in the 3$^{rd}$ dynamic function call\footnote{Note that the dynamic function indices
start from 0.} for PR, due to the significant
difference in the noise tolerance of 3$^{rd}$ and 4$^{th}$ calls (which is not
the case under DPS, as Fig.~\ref{fig:pol_pr}(a) reveals).
A similar matching, although less visible, applies for PF, in 
Fig.~\ref{fig:pol_pf}.

In summary,  we observe that DPS+ can effectively bound the accuracy loss,
although overshoots still apply for HS under DPS+.
To handle these cases, DPS+ can be refined to track the actual call graphs of data
dependent dynamic functions for precision matching. Such refinement is also
likely to untap even more opportunities, considering the cases in
Tables~\ref{tab:final_error},~\ref{tab:final_error_alg2} with a large gap between the
actual accuracy loss and {\em targetAccLoss}.

\begin{figure}[h]
  \begin{center}
	  \includegraphics[width=0.5\columnwidth]{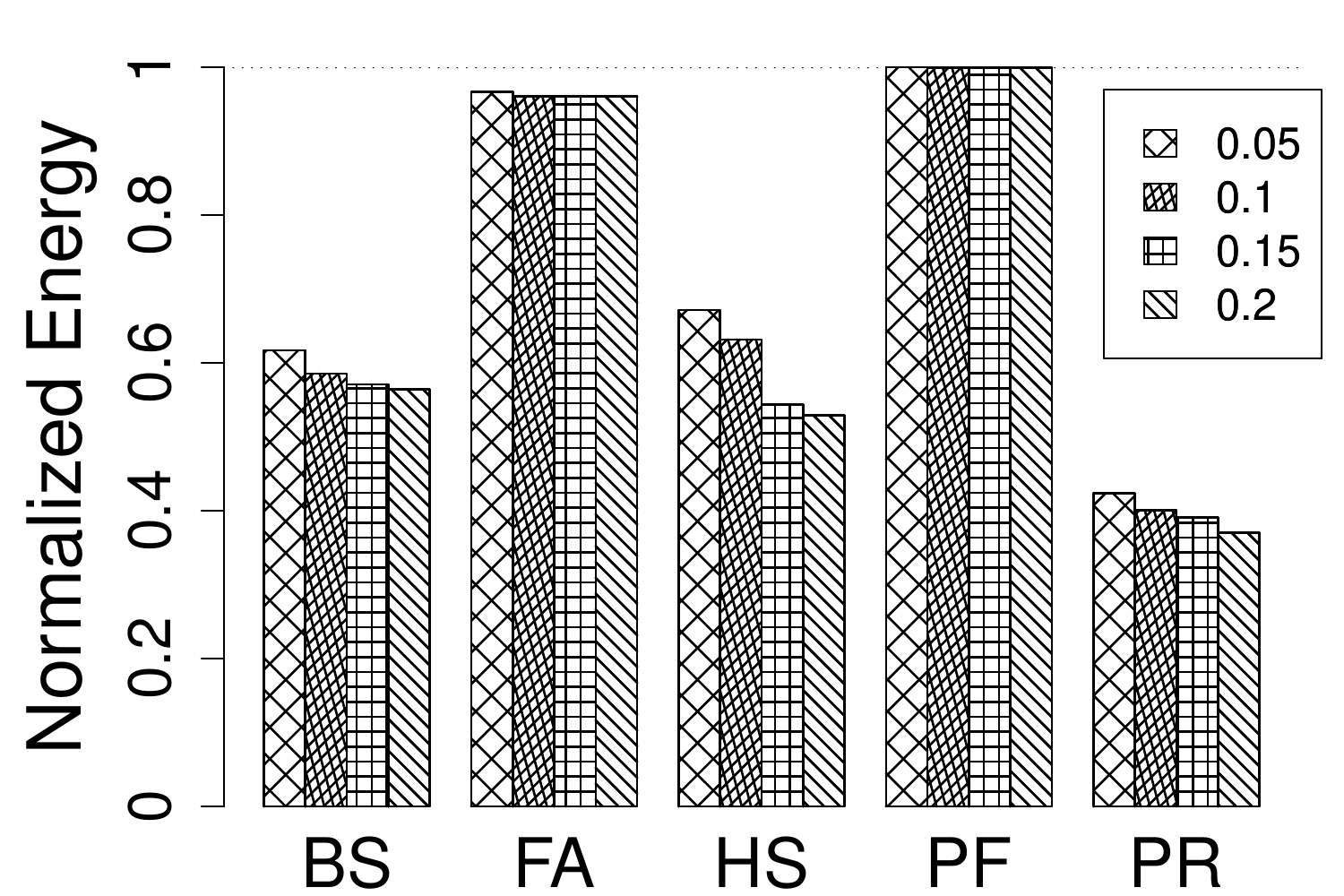}
	\caption{Energy consumption under DPS for different {\em targetAccLoss} values.
	  \label{fig:energyDPS}}
  \end{center}
\end{figure}

Fig.~\ref{fig:energyDPS} captures energy savings under DPS for {\em targetAccLoss} values
between 0.05 and 0.2 
at increments of 0.05.
Savings span 
0.09\% to 63.8\%. 
Notable energy savings apply to BS, HS, and PR. 
Savings for FA and PF are modest.
DPS(+) excludes floating point
instructions from shared libraries, which likely hurts both of these
applications.
At the same time, for PF, the ratio of
$\#instructions_{C,no}/\#instructions_{C,o}$ is around 10, which severely limits energy
savings independent of {\em targetAccLoss} value. 
Also, the most energy-hungry dynamic functions feature the lowest number
of omitted mantissa bits in Fig.~\ref{fig:pol_pf}.

DPS+ renders less (or at most equal) number of omitted
bits when compared to DPS, hence may leave potential energy savings untapped in
trying to limit the accuracy loss.
However, we find that the maximum difference between energy savings of DPS
and DPS+ barely reaches 4.7\%. On average, the difference remains around 1.44\%.

\redHL{
   Figure~\ref{fig:energyMinDPS} captures the energy profile under SPS+.
   Overall, SPS+
   renders a higher energy consumption than DPS(+). 
BS is not the only application which
loses the energy benefits of {\em dynamic} precision scaling under SPS+.
For example, energy consumption
of PR increases by 45\% when compared to DPS. Overall, the increase in energy
consumption varies between 1\% to 78\% when we consider all applications.}

\begin{figure}[h]
  \begin{center}
        \subfloat[DPS]{
          \includegraphics[width=0.5\columnwidth]{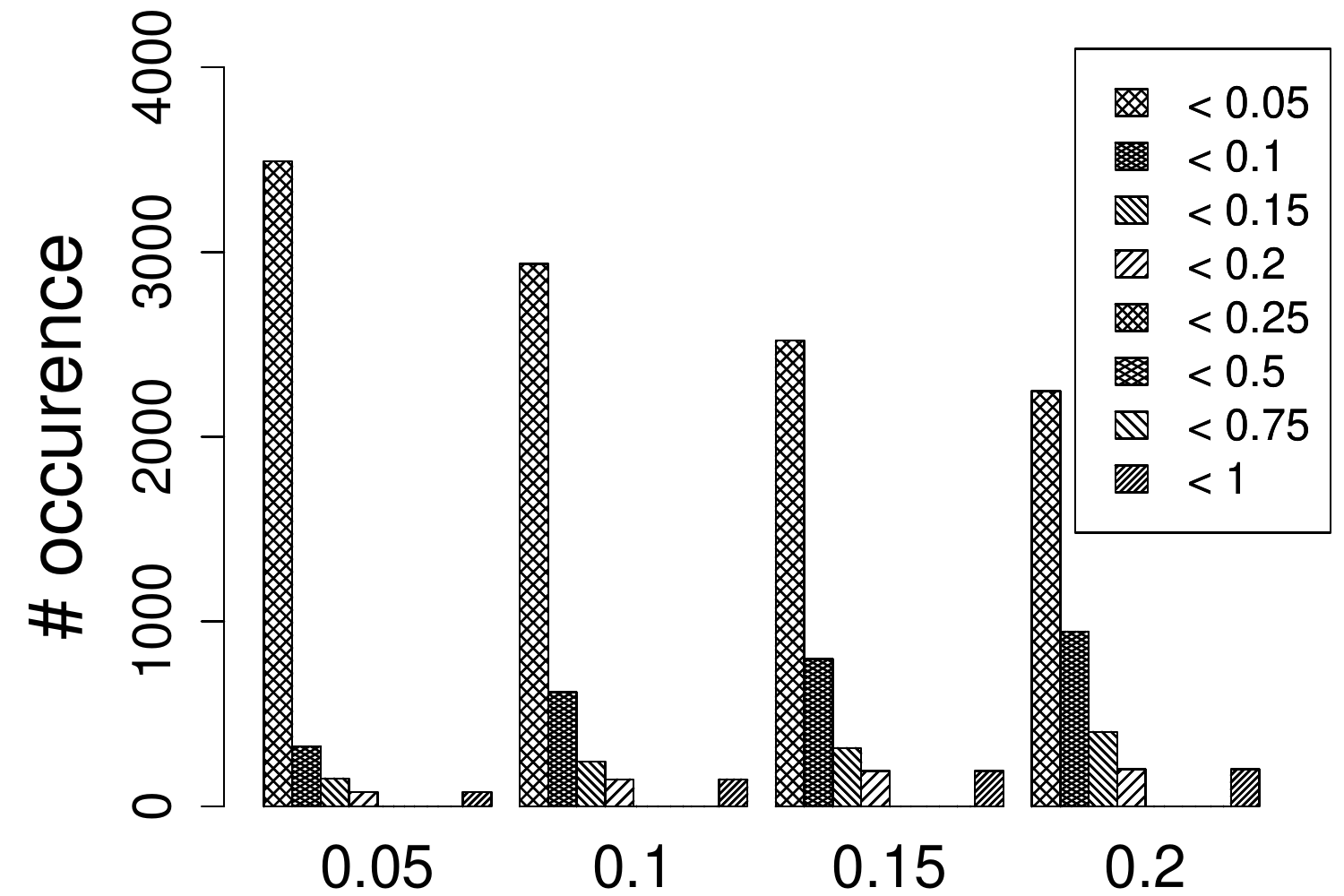}
    	}
        \subfloat[DPS+]{
          \includegraphics[width=0.5\columnwidth]{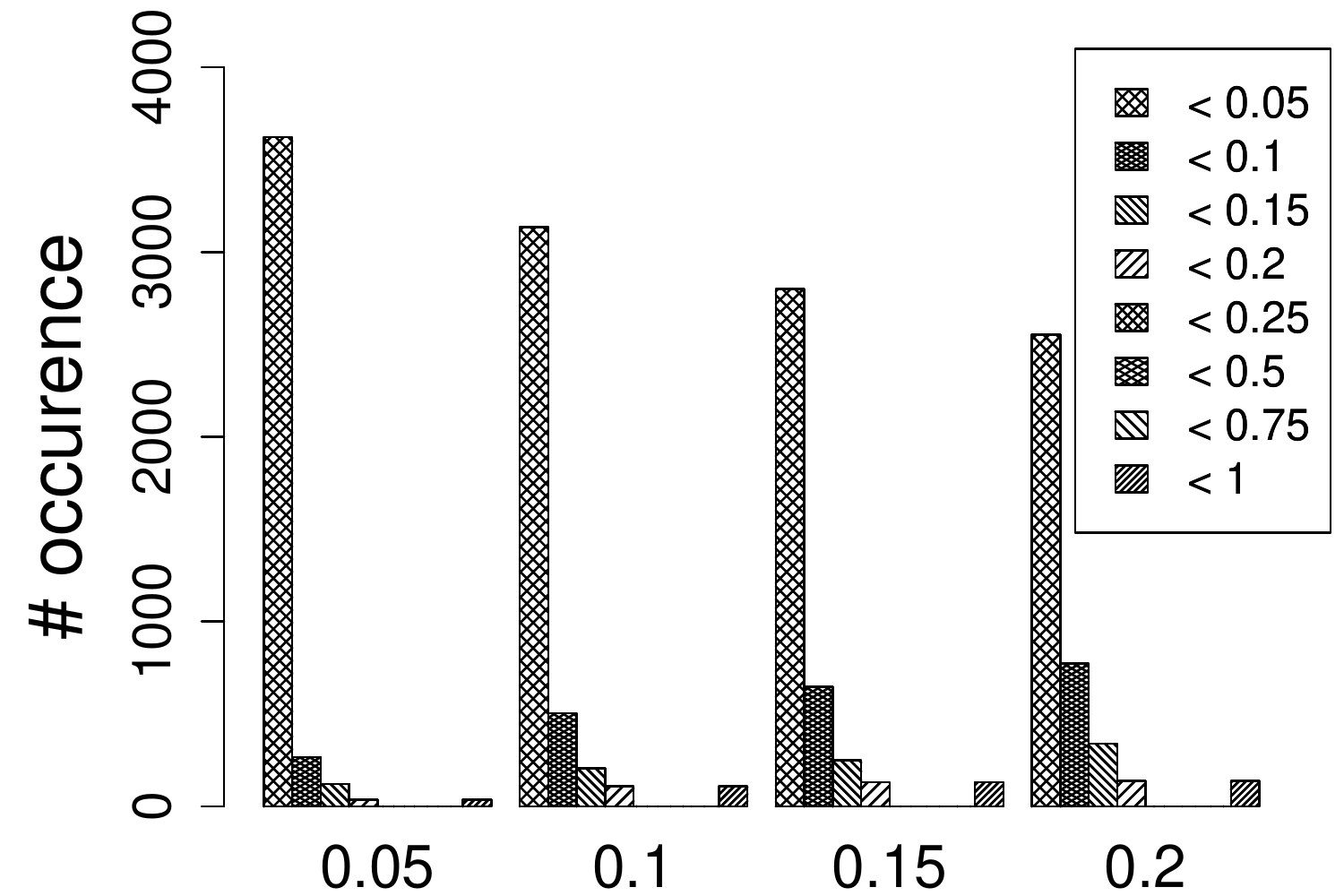}
    	}
		\caption{\redHL{Accuracy loss distribution for BS.
	  \label{fig:hist_bs}}}
  \end{center}
\end{figure}

\begin{figure}[h]
  \begin{center}
        \subfloat[DPS]{
          \includegraphics[width=0.5\columnwidth]{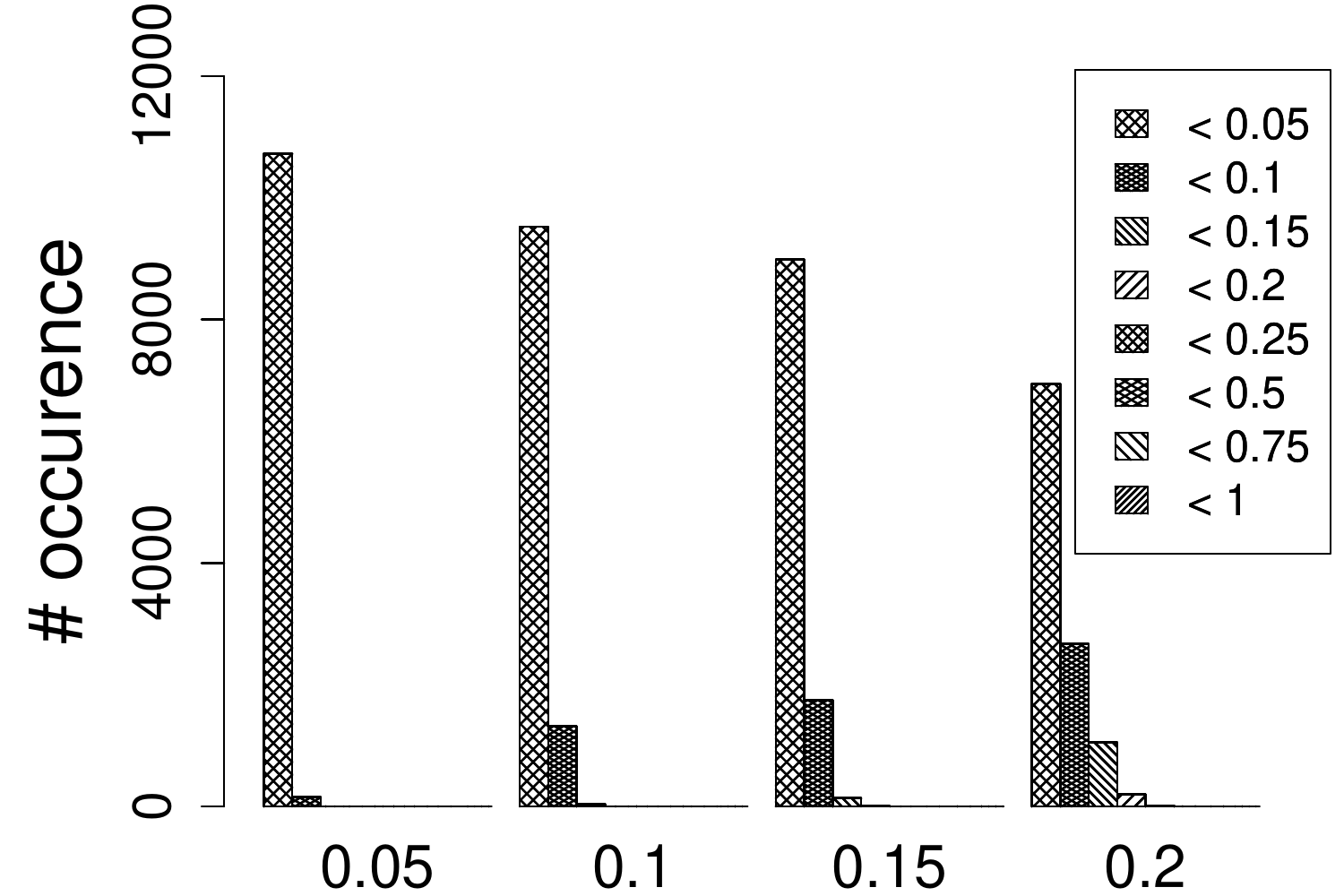}
        }
        \subfloat[DPS+]{
          \includegraphics[width=0.5\columnwidth]{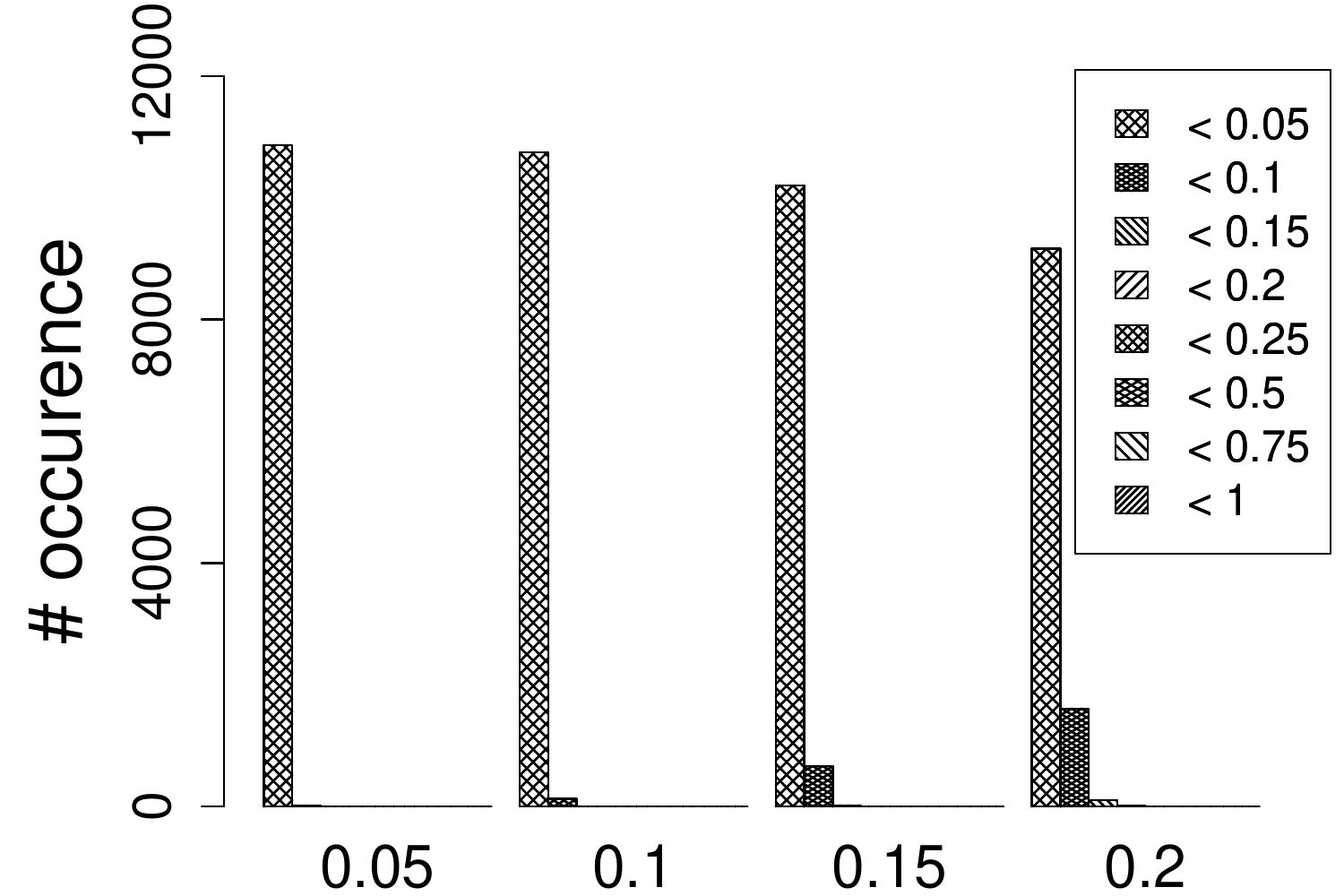}
        }
                \caption{\redHL{Accuracy loss distribution for PR with gnu04 input set.
          \label{fig:hist_pr_gnu04}}}
  \end{center}
\end{figure}

\begin{figure}[h]
  \begin{center}
        \subfloat[DPS]{
          \includegraphics[width=0.5\columnwidth]{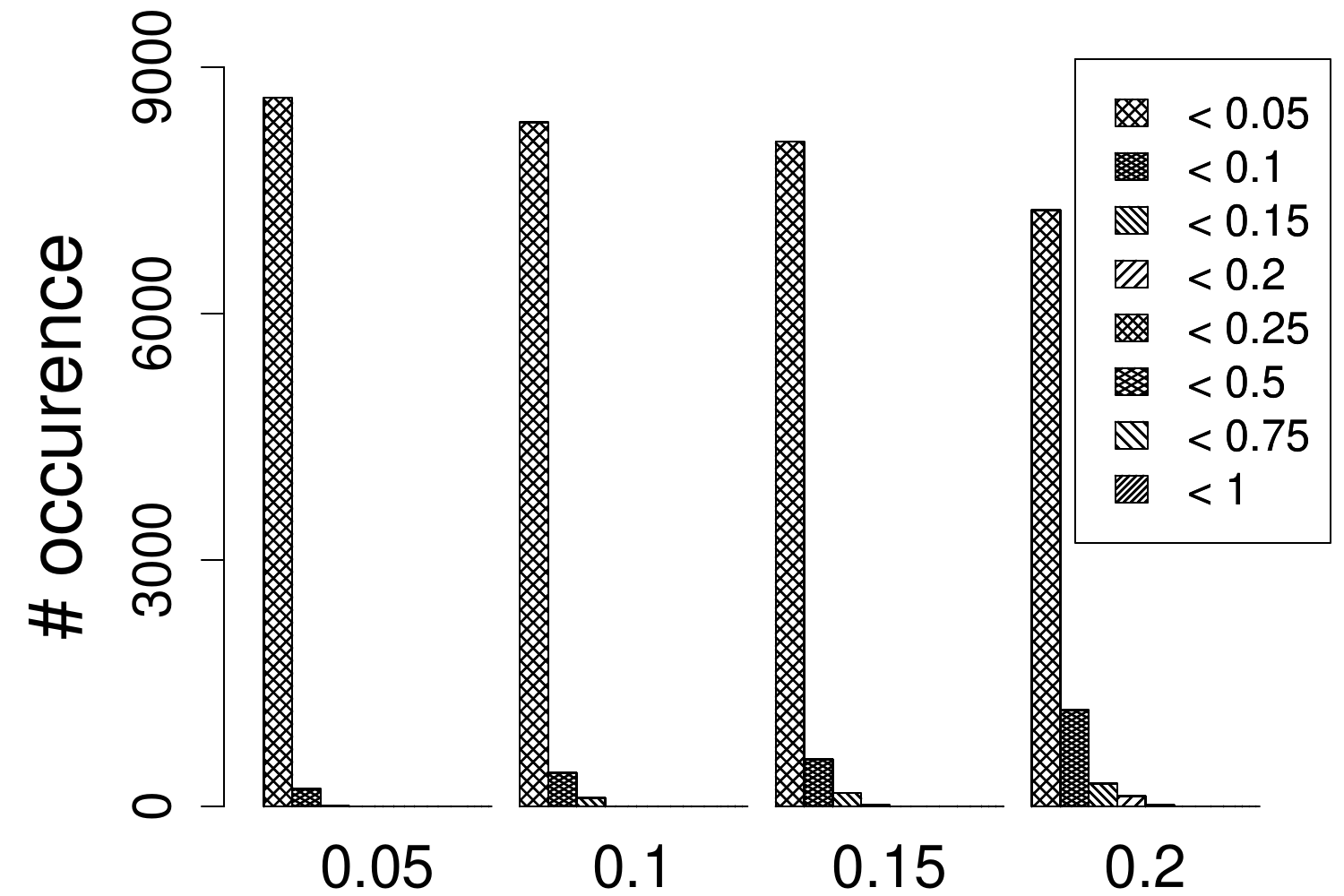}
        }
        \subfloat[DPS+]{
          \includegraphics[width=0.5\columnwidth]{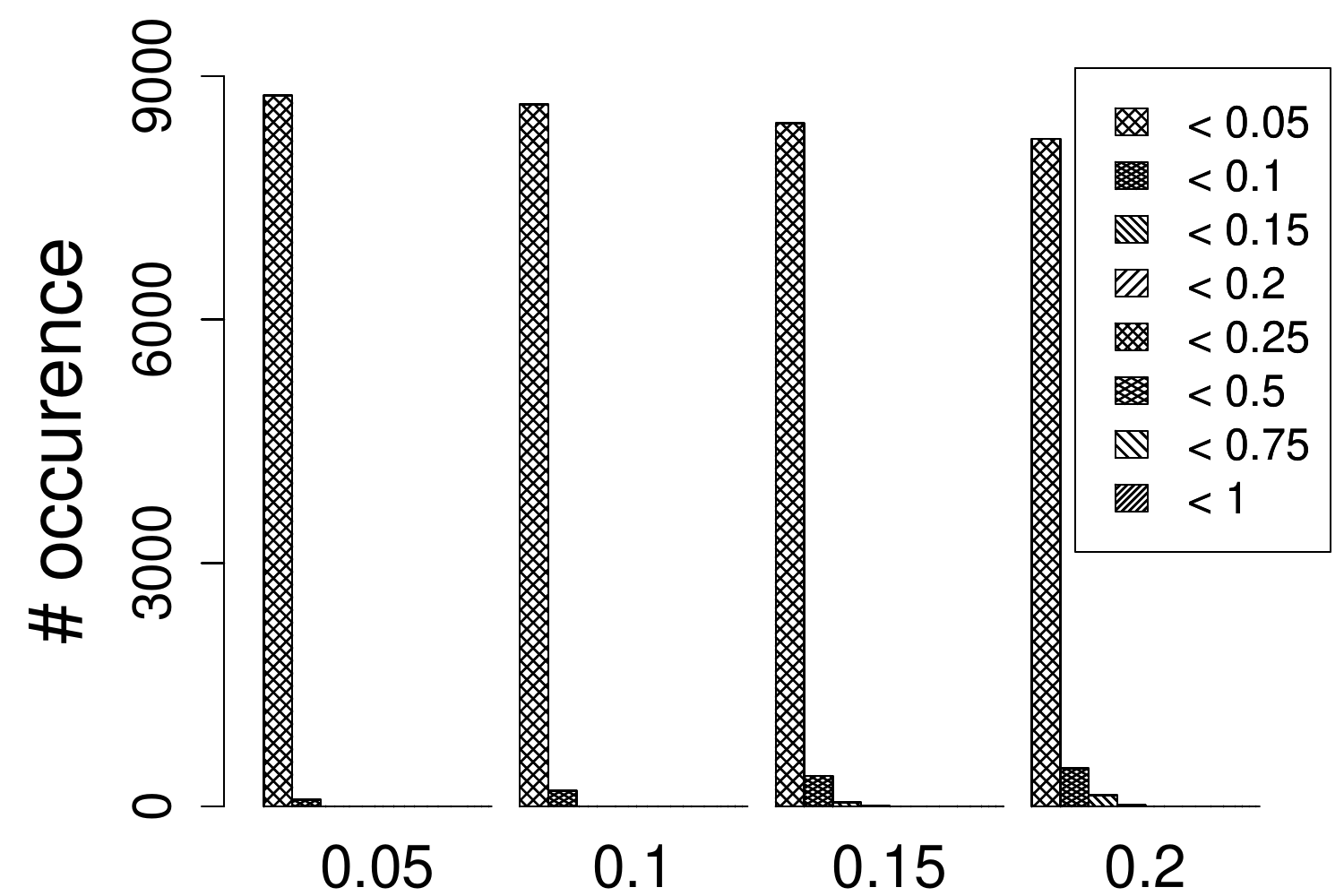}
        }
                \caption{\redHL{Accuracy loss distribution for PR with gnu05 input set.
          \label{fig:hist_pr_gnu05}}}
  \end{center}
\end{figure}

\begin{figure}[h]
  \begin{center}
        \subfloat[DPS]{
          \includegraphics[width=0.5\columnwidth]{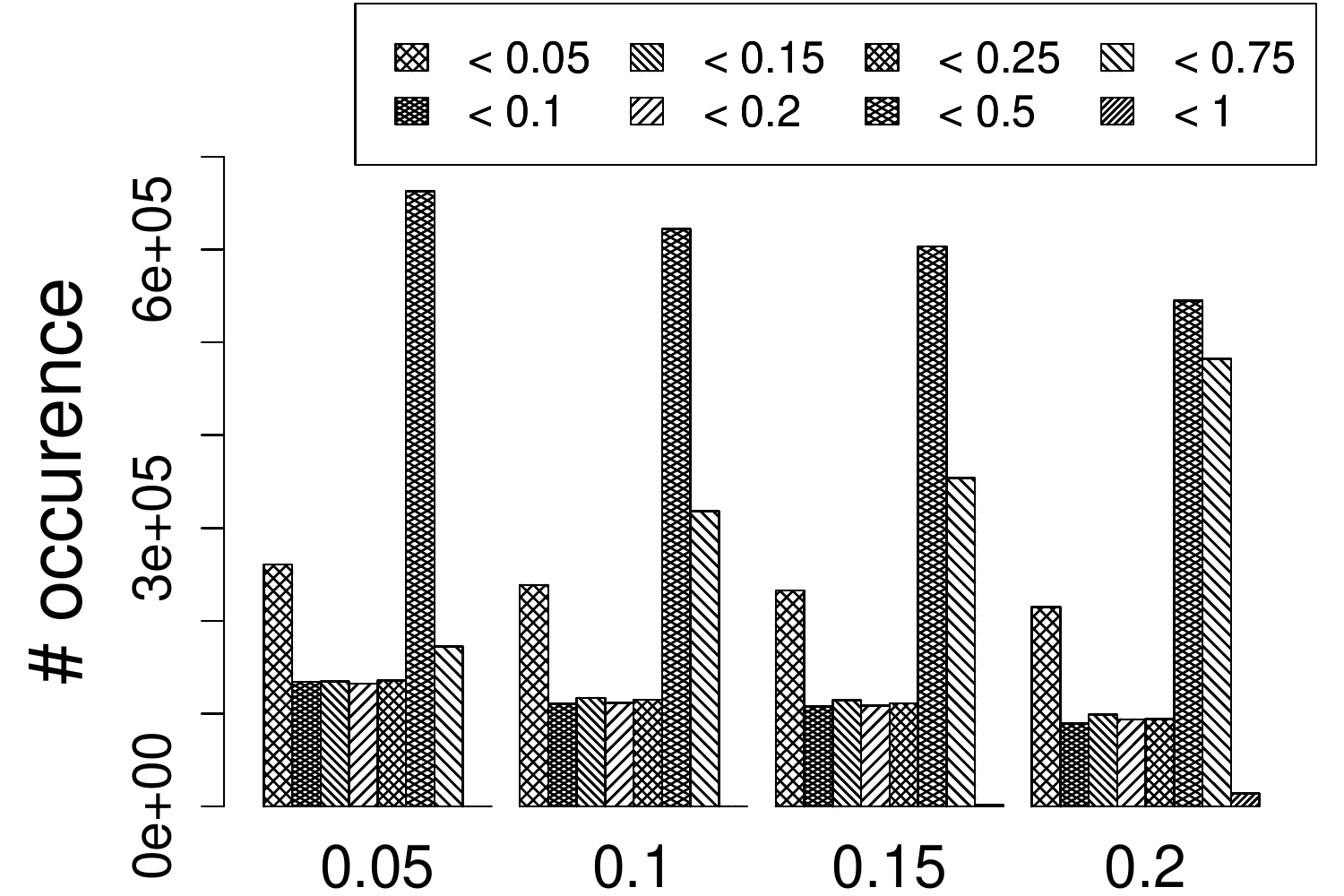}
        }
        \subfloat[DPS+]{
          \includegraphics[width=0.5\columnwidth]{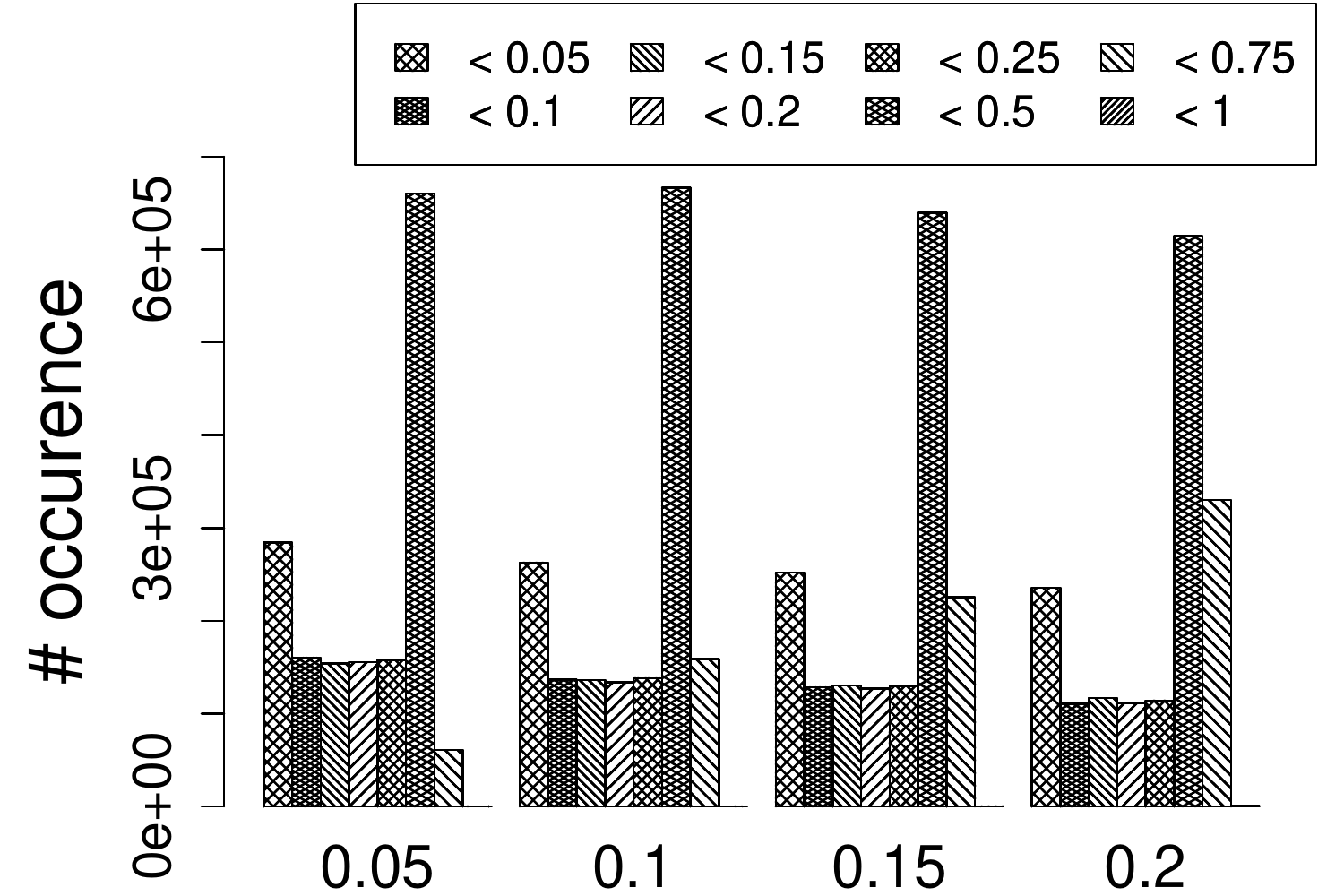}
        }
                \caption{\redHL{Accuracy loss distribution for PR with pk input set.
          \label{fig:hist_pr_pk}}}
  \end{center}
\end{figure}

\noindent \redHL{
{\bf A note on accuracy metric:} In quantifying the accuracy loss, we
used mean relative error (i.e., mean relative accuracy loss) as a generic accuracy metric, as explained in
Section~\ref{sec:setup}. Mean relative error may be misleading if the
standard deviation assumes a large value. To quantify the standard deviation,
Figures~\ref{fig:hist_bs}--\ref{fig:hist_pr_pk} depict the distribution of
component accuracy loss values (as captured by the legends). The x-axis captures
different values of {\em
targetAccLoss}. 
For example, for BS, we observe that 
84\% to 92\% of relative accuracy loss values fall below
the given {\em
targetAccLoss} under DPS; and 92\% to 98\%, under DPS+. A similar trend
holds for PR, considering different inputs. 
}

\subsection{Input Sensitivity}

\begin{table}[ht]
\centering
\caption{\redHL{Input Sensitivity Analysis for Pagerank (PR) with different datasets from
the Snap Database\protect\cite{snapnets}. Only \textbf{gnu04} is deployed for
profiling.}}
\label{tab:pr_sens}
\begin{tabular}{|l|l|l|l|l|l|}
\hline
                                &                   & \textbf{0.05} & \textbf{0.1}  & \textbf{0.15} & \textbf{0.2}  \\ \hline
\multirow{3}{*}{\textbf{gnu04}} & \textbf{DPS}      & 0.0849 & 0.1416  & 0.1714  & 0.2582  \\ \cline{2-6} 
                                & \textbf{DPS+}     & 0.0758 & 0.0848 & 0.1428 & 0.1646  \\ \cline{2-6} 
                                & \textbf{SPS+} & 0.0615 & 0.0615 & 0.1308 & 0.1308 \\ \hline
\multirow{3}{*}{\textbf{gnu05}} & \textbf{DPS}      & 0.1029 & 0.1499 & 0.1805 & 0.2423 \\ \cline{2-6} 
                                & \textbf{DPS+}     & 0.0846 & 0.0988 & 0.1597 & 0.1763 \\ \cline{2-6} 
                                & \textbf{SPS+} & 0.0764 & 0.0764 & 0.1406 & 0.1406 \\ \hline
\multirow{3}{*}{\textbf{gnu08}} & \textbf{DPS}      & 0.1906 & 0.2567 & 0.3232 & 0.3900 \\ \cline{2-6} 
                                & \textbf{DPS+}     & 0.1604 & 0.1857 & 0.2829 & 0.3132 \\ \cline{2-6} 
                                & \textbf{SPS+} & 0.1400 & 0.1400 & 0.2463 & 0.2463 \\ \hline
\multirow{3}{*}{\textbf{gnu25}} & \textbf{DPS}      & 0.0395 & 0.0833 & 0.0986 & 0.1730 \\ \cline{2-6} 
                                & \textbf{DPS+}     & 0.0276 & 0.0391 & 0.0650 & 0.0896 \\ \cline{2-6} 
                                & \textbf{SPS+} & 0.0244 & 0.0244 & 0.0525 & 0.0525 \\ \hline
\multirow{3}{*}{\textbf{pk}}    & \textbf{DPS}      & 0.9107 & 0.9211 & 0.9490 & 0.9524 \\ \cline{2-6} 
                                & \textbf{DPS+}     & 0.9050 & 0.9101 & 0.9452 & 0.9488 \\ \cline{2-6} 
                                & \textbf{SPS+} & 0.8832 & 0.8832 & 0.9317 & 0.9317 \\ \hline
\multirow{3}{*}{\textbf{wg}}    & \textbf{DPS}      & 0.9310 & 0.9572 & 0.9611 & 0.9787 \\ \cline{2-6} 
                                & \textbf{DPS+}     & 0.9019 & 0.9269 & 0.9383 & 0.9508 \\ \cline{2-6} 
                                & \textbf{SPS+} & 0.7813 & 0.7813 & 0.8839 & 0.8839 \\ \hline
\end{tabular}
\end{table}

\redHL{In order to quantify the input dependence due to profiling, we  
experiment with
PR, which features a rich set of inputs.
As other profiling based
approaches, DPS(+) is input dependent. However, the degree of this dependence
changes from application to application.
At the same time, when the properties (such as size and value distribution) of two input datasets are close to each other, 
{\em \#omittedBits} per dynamic call, as identified by profiling under one
dataset, may render reasonable output when applied to execution under another
dataset.
Table~\ref{tab:pr_sens} quantifies this effect for PR. This table is the
equivalent of Tables~\ref{tab:final_error}--~\ref{tab:final_error_minDPS},
except that only the \textbf{gnu04} input dataset is used for profiling. In
other words, {\em \#omittedBits} as determined by a profiling pass for
\textbf{gnu04} is imposed when running the same application with different
input datasets (as tabulated in the first column of Table~\ref{tab:pr_sens}). 
The number of vertices
in gnuXX graphs vary between 6K to 10K (and reaches 22K for gnu25); the number
of edges, between 20K
and 40K (and reaches 54K for gnu25).
For these inputs, PR features a relatively  
{weak input dependence}. 
We experiment with two more graphs:
soc-pokec (pk) and web-Google (wg) from the same graph database. These
graphs have 1600K and 800K vertices, and 30M and 5M edges, respectively. As Table
\ref{tab:pr_sens} captures, the picture changes for these graphs, and the
discrepancy becomes notable.}

\redHL{On the other hand, for some applications (such as BS and HS), both
the number and the input-sensitivity of dynamic function calls strongly depend
on input size and value distribution. Any
change in inputs in this case is likely to cause a notable discrepancy between
the profiled and actual execution outcomes. 
}

\section{Related Work}
\label{sec:rel}
Static precision reduction for floating point arithmetic has been heavily
studied (e.g.,~\cite{rutenbar}).  Adaptive precision reduction, on the other
hand, has been explored in the context of physics simulation to minimize the
area cost of floating point units (FPUs)~\cite{yeh2007} and for digital signal
processing~\cite{dspDPS}.  In contrast, our focus is to show the opportunity in
general purpose computing.  Our study evaluates DPS for floating point
approximation, however, DPS can be applied to the integer data path, as well.
At the same time, our goal is boosting the power efficiency considering a
broader emerging class of RMS applications~\cite{rms}.  Automated program
analysis tools to help developers tune floating point
precision~\cite{preci,linderman} fit well into the {\em offline profiling} stage
of DPS, but the existing body of work in this domain usually does not explore
adaptive precision tuning at runtime.

One end of the spectrum for detection of approximation in software is
EnerJ~\cite{enerJ}.  In this work, the authors do not automate the process and
require programmers to define approximate data types.  On the other hand,
Chisel~\cite{chisel} provides a semi-automated approach which tries to maximize
both accuracy and energy efficiency at the same time.  However, Chisel still
requires the programmer to specify approximate program segments and probability
that the specified function should execute correctly.  In our approach, we are
trying to detect possible noise tolerant phases automatically.  Proposed
methodology in this work can be used to detect and reduce the time spend to tune
the approximate code segments for ~\cite{enerJ} and ~\cite{chisel}.

As oppose to semi or non-automated mechanisms, SAGE~\cite{sage} proposes an
automated approach by using online monitoring mechanism designed for GPU
kernels.  Closest to our work, Approxilyzer~\cite{approxilyzer} tries to find
noise tolerant instructions in an application and classifies them as Masked,
SDC-Good/Maybe/Bad, and detectable errors.  This approach is automated; however,
it is only limited to single bit errors.  In this work, we expand the code
region and work on the function granularity to enable more energy reduction. 

In addition to the holistic approximate computing approaches, configurable
approximate floating point arithmetic units attracted significant attention
~\cite{fpmul1}, ~\cite{fpmul2}.

\section{Conclusion \& Discussion}
\label{sec:conc}
This paper provides a proof-of-concept analysis for dynamic precision scaling,
DPS, which tailors the arithmetic precision to changes in the application's
noise tolerance within the course of execution.  As a case study, \redHL{without
loss of generality}, we confine our analysis to the floating point data path.
However, DPS can also cover integer operations, where the main complication
comes from identification, and hence,  exclusion of memory address calculations
(i.e., pointer arithmetic) from approximation.  

We envision a practical DPS   implementation to comprise three basic modules:
(i) an {\em offline profiler} to identify and demarcate application phases of
different noise tolerance characteristics; (ii) a {\em runtime monitor} to track
temporal changes in workload's noise tolerance, and (iii) an {\em accuracy
controller} to adjust the arithmetic precision on the fly accordingly.  The
differences in the design of these three modules give rise to different points
in the DPS design space. 

The {\em offline profiler} and {\em runtime monitor} should be able to capture
fine-grain temporal changes in application's noise tolerance. As the noise
tolerance of RMS applications is mainly algorithmic, software intervention is
inevitable -- e.g., in the form of code annotations to demarcate varying degrees
of noise tolerance.  To communicate such annotations to the hardware,
programming language extensions~\cite{enerJ} may help. At the same time, noise
tolerance is input-dependent, rendering profiling based approaches such
as~\cite{rumba,debugApprox} (including ours) necessary. The ideal solution may
be hidden in -- yet to be explored -- correlations between hardware-observable
features (similar to performance counter outcome) and noise tolerance at the
application level.

The {\em accuracy controller} design space spans software, hardware, or hybrid
implementations -- similar to DVFS controllers. For example, if the processor
features functional units of reduced precision, the controller is in charge of
scheduling (more) noise-tolerant phases to lower-precision arithmetic units. The
processor may also accommodate functional units of reconfigurable precision in
order to harvest power efficiency under DPS.  In each case, a very stringent
budget applies for the power and performance overhead of the accuracy
controller.  

Finally, we should also note that {\em tailoring the degree of approximation to
changes in the application's noise tolerance within the course of execution} is
a generic paradigm which can be adapted to many other approximation techniques
beyond precision scaling. This paper can be regarded as a case study in this
respect, as well.



\end{document}